\documentclass[acmtog]{acmart}

\acmConference[SIGGRAPH '26 Conference Papers]{Special Interest Group on Computer Graphics and Interactive Techniques Conference Conference Papers}{July 19--23, 2026}{Los Angeles, CA, USA}

\acmBooktitle{SIGGRAPH Conference Papers '26, July 19--23, 2026, Los Angeles, CA, USA}

\acmISBN{979-8-4007-2554-8/26/07}
\acmDOI{https://doi.org/10.1145/3799902.3811207}

\usepackage{booktabs} 

\usepackage[ruled]{algorithm2e} 
\usepackage{subcaption}
\usepackage{enumitem}
\usepackage{multirow}

\SetAlFnt{\small}
\SetAlCapFnt{\small}
\SetAlCapNameFnt{\small}
\SetAlCapHSkip{0pt}
\begin{document}
\title{BrepForge: Factorized B-rep Synthesis via Wireframe Composition and Boundary-Conditioned Surface Instantiation}

\author{Jing Li}
\orcid{0009-0007-5792-3796}
\affiliation{%
 \institution{University of Science and Technology of China}
 \country{China}}
\email{jingli99@mail.ustc.edu.cn}

\author{Yihang Fu}
\orcid{0009-0004-8621-652X}
\affiliation{%
 \institution{University of Science and Technology of China}
 \country{China}}
\email{fuyihang@mail.ustc.edu.cn}

\author{Falai Chen}
\orcid{0000-0002-9898-5922}
\affiliation{%
 \institution{University of Science and Technology of China}
 \streetaddress{104 Jamestown Rd}
 \country{China}}
\email{chenfl@ustc.edu.cn}



\begin{abstract}
Boundary representation (B-rep) is the de facto standard for modern CAD, yet learning-based B-rep synthesis remains challenging due to the tight coupling between discrete topology and continuous geometry. We observe a fundamental asymmetry in B-reps: while wireframe composition involves high-entropy structural decisions, the interior surface geometry is largely constrained by its boundary loops. Motivated by this observation, we propose BrepForge, a generative framework that factorizes B-rep synthesis into two stages: wireframe composition and boundary-conditioned surface instantiation. In the first stage, a face-aware autoregressive model serializes the wireframe into structured sequences that explicitly encode hierarchical Vertex–Edge–Face (V–E–F) connectivity, yielding a topologically complete scaffold. In the second stage, precise surface geometries are instantiated by incorporating learning-free geometric priors derived from boundaries, transforming the complex synthesis task into a structured refinement process. This factorized approach ensures both topological integrity and geometric precision, effectively addressing the inherent complexities of B-rep modeling. Extensive experiments demonstrate that BrepForge outperforms existing baselines with superior geometric complexity and topological validity.
\end{abstract}

%
%

\ccsdesc[500]{Applied computing~Computer-aided design}
\ccsdesc[300]{Computing methodologies~Shape Modeling}

%
%

\keywords{Boundary representation, Wireframe, Generative model}

\maketitle


\section{Introduction}
Boundary representation (B-rep) \cite{weiler1986topological} is the standard for modern computer-aided design (CAD), offering mathematically exact descriptions of solid geometry. Unlike discrete meshes or point clouds, B-rep defines shapes through the explicit coupling of continuous geometric primitives—vertices, curves, and surfaces—with a topological hierarchy. This integration enables high-fidelity modeling of sharp features and smooth surfaces, preserving essential structural information for downstream tasks like parametric editing and precision manufacturing.

Despite its importance, learning-based B-rep synthesis remains challenging due to the need to simultaneously capture discrete topology and continuous geometry. Command-based methods \cite{wu2021deepcad, xu2022skexgen, xu2023hierarchical, khan2024text2cad, guo2025cadtrans} address this problem indirectly by predicting sequences of modeling operations. However, these approaches are typically limited to simple ``sketch-and-extrude'' patterns and depend on scarce datasets that provide complete modeling histories \cite{willis2021fusion, wu2021deepcad}. Alternatively, direct synthesis methods offer greater flexibility but face a fundamental trade-off. Early multi-stage frameworks \cite{jayaraman2022solidgen, xu2024brepgen, li2025dtgbrepgen, li2025stitch} decompose B-rep generation into cascades of sub-networks that predict individual primitives. As a result, they are prone to cascading error accumulation, where small inaccuracies in early stages can lead to invalid final B-reps. More recent single-stage approaches aim to learn B-reps end-to-end, either as unified graphs \cite{guo2025brepgiff, lai2025graphbrep} or as exhaustive one-dimensional sequences \cite{liu2025hola, li2025brepgpt, xu2025autobrep}. However, compressing all geometric entities and their interdependencies into a single representation imposes a substantial learning burden. Such ``all-in-one'' strategies often encounter training instability and struggle to scale to complex models, where the high density of structural elements makes it difficult for a single model to maintain global coherence.

In this work, we revisit B-rep structures and identify a critical asymmetry: the wireframe carries a disproportionately large share of a solid’s information compared to its faces. This asymmetry arises from strong inductive biases in CAD design, where most faces are instantiated from low-degree canonical primitives (e.g., planes and cylinders). Because the interiors of such primitives are effectively determined by their boundaries—and even free-form surfaces (e.g., NURBS patches) are heavily constrained by their edge scaffolds—the space of admissible surfaces becomes highly restricted once the wireframe is specified. This imbalance implies that wireframe composition entails high-entropy structural reasoning, whereas surface generation largely amounts to refining an already constrained scaffold. Consequently, modeling the entire B-rep in a monolithic manner is redundant and unnecessarily burdens the learning process.  

Motivated by this observation, we propose BrepForge, a factorized framework that decouples B-rep synthesis into wireframe composition and boundary-conditioned surface instantiation. In the first stage, BrepForge synthesizes a complete wireframe that specifies both geometry and hierarchical vertex–edge–face (V–E–F) connectivity. Unlike prior wireframe generators that produce unstructured edge collections \cite{ma2024generating, xue2024neat, ma2025clr, zhu2023nerve}—which lack the explicit loop-to-face associations required for solid modeling—we introduce a face-aware autoregressive formulation. By serializing the wireframe into structured sequences in which vertices and edges are interleaved within ordered loops, the model directly encodes topological constraints among B-rep entities. As a result, the generated scaffold is not merely a set of geometric curves, but a topologically closed foundation suitable for subsequent surface instantiation. In the second stage, BrepForge populates this scaffold with precise surface geometries. Leveraging the observation that boundary representations strongly constrain surface interiors, we incorporate learning-free geometric priors to recast surface synthesis as a structured refinement process. To maintain global coherence across all faces, a Transformer-based architecture \cite{vaswani2017attention} facilitates information exchange at the inter-face level. This hybrid strategy substantially reduces the learning burden while preserving high-fidelity geometric accuracy.

The main contributions of this work are as follows:
\begin{itemize}[leftmargin=2em]
\item We propose a factorized generative framework for B-rep synthesis that decomposes the task into wireframe composition and surface instantiation, alleviating the representation bottleneck of monolithic approaches.
\item We introduce a face-aware sequential encoding for B-rep wireframes that jointly models geometry and hierarchical Vertex-Edge-Face (V-E-F) connectivity, enabling the generation of topologically closed wireframe scaffolds.
\item We design a boundary-conditioned surface instantiation module that incorporates analytical geometric priors derived from the wireframe, recasting surface synthesis as a structured refinement process.
\end{itemize}

\section{Related Work}
\label{sec: related work}

\subsection{CAD Command Generation}
Command-based methods synthesize CAD solids as sequences of parametric modeling operations, explicitly emulating traditional procedural design workflows. DeepCAD \cite{wu2021deepcad} pioneered this paradigm with a Transformer-based autoencoder, while subsequent works such as SkexGen \cite{xu2022skexgen} and HNC \cite{xu2023hierarchical} introduced hierarchical representations to better capture long-range dependencies. Recently, this line of research has expanded toward multimodal and conditional settings, leveraging large language models and vision–language frameworks to enable CAD synthesis from text, images, and point clouds \cite{khan2024text2cad, xu2024cad, li2025cad}. Parallel efforts have explored interactive design via text-based editing \cite{zhang2024flexcad, yuan2025cad} and improved operation sequencing through reinforcement learning \cite{deng2025mamtiff, yin2025rlcad}. Despite these advances, command-based approaches remain largely restricted to simple sketch-and-extrude operations and fundamentally rely on scarce construction histories, which are significantly less available than large-scale B-rep datasets such as ABC \cite{koch2019abc}.

\subsection{3D Wireframe Generation}
3D wireframes serve as a skeletal abstraction of CAD geometry, capturing the essential connectivity of shapes. Early studies primarily focused on rectilinear segments \cite{huang2024pbwr, ma2024generating}, extracting junctions and lines from images \cite{ma20223d, pautrat2023deeplsd} or point clouds \cite{liu2021pc2wf, tan2022coarse}. To capture the curved profiles prevalent in engineering, subsequent research transitioned toward curve-based paradigms, utilizing parameterized Bézier curves or B-splines \cite{ye2023nef, wang2022neural}. Complementary approaches further leverage implicit neural representations and distance fields \cite{guo2022complexgen, matveev2022def, cherenkova2023sepicnet, matveev20213d} to improve geometric fitting fidelity. However, existing methods predominantly focus on low-level Vertex–Edge (V–E) connectivity. While recovering a plausible skeletal graph of the shape, they do not model the hierarchical Vertex–Edge–Face (V–E–F) topology required for solid modeling. Consequently, these outputs lack explicit face–loop associations, precluding their direct instantiation into valid solid models.

\subsection{Direct B-rep Generation}
Direct B-rep synthesis models complete boundary representations without requiring construction histories. SolidGen \cite{jayaraman2022solidgen} pioneered this domain by employing a Transformer-based framework with a pointer network \cite{vinyals2015pointer} to sequentially predict vertices, edges, faces, and their topological relationships. Subsequently, BrepGen \cite{xu2024brepgen} utilized a suite of diffusion models \cite{ho2020denoising, song2020denoising} for progressive geometric generation, while DTGBrepGen \cite{li2025dtgbrepgen} further decoupled topology and geometry into separate modules to enforce topological validity. However, these multi-stage pipelines often suffer from cumulative error propagation, where early-stage inaccuracies lead to severe geometric inconsistencies. To alleviate this issue, recent work has explored more unified generation strategies. BrepDiff \cite{lee2025brepdiff} adopts a single diffusion model to generate face geometries, but still relies on post-processing steps such as surface intersection to recover valid topology. More integrated approaches such as HoLa-Brep \cite{liu2025hola} and BrepGiff \cite{guo2025brepgiff} attempt to encode the entire B-rep into a unified latent space or a global topological graph. Inspired by the success of GPT-style models in language and vision \cite{brown2020language, chen2020generative, ramesh2021zero, siddiqui2024meshgpt}, concurrent works such as AutoBrep \cite{xu2025autobrep} and BrepGPT \cite{li2025brepgpt} serialize B-reps into exhaustive 1D sequences for autoregressive generation. While these monolithic approaches avoid explicit stage-wise decomposition, they impose a substantial learning burden due to long sequence lengths and dense structural dependencies, making it difficult to maintain global coherence for complex industrial parts.

\begin{figure*}[tb]
\captionsetup{skip=4pt}
  \centering
  \includegraphics[width=0.8\linewidth]{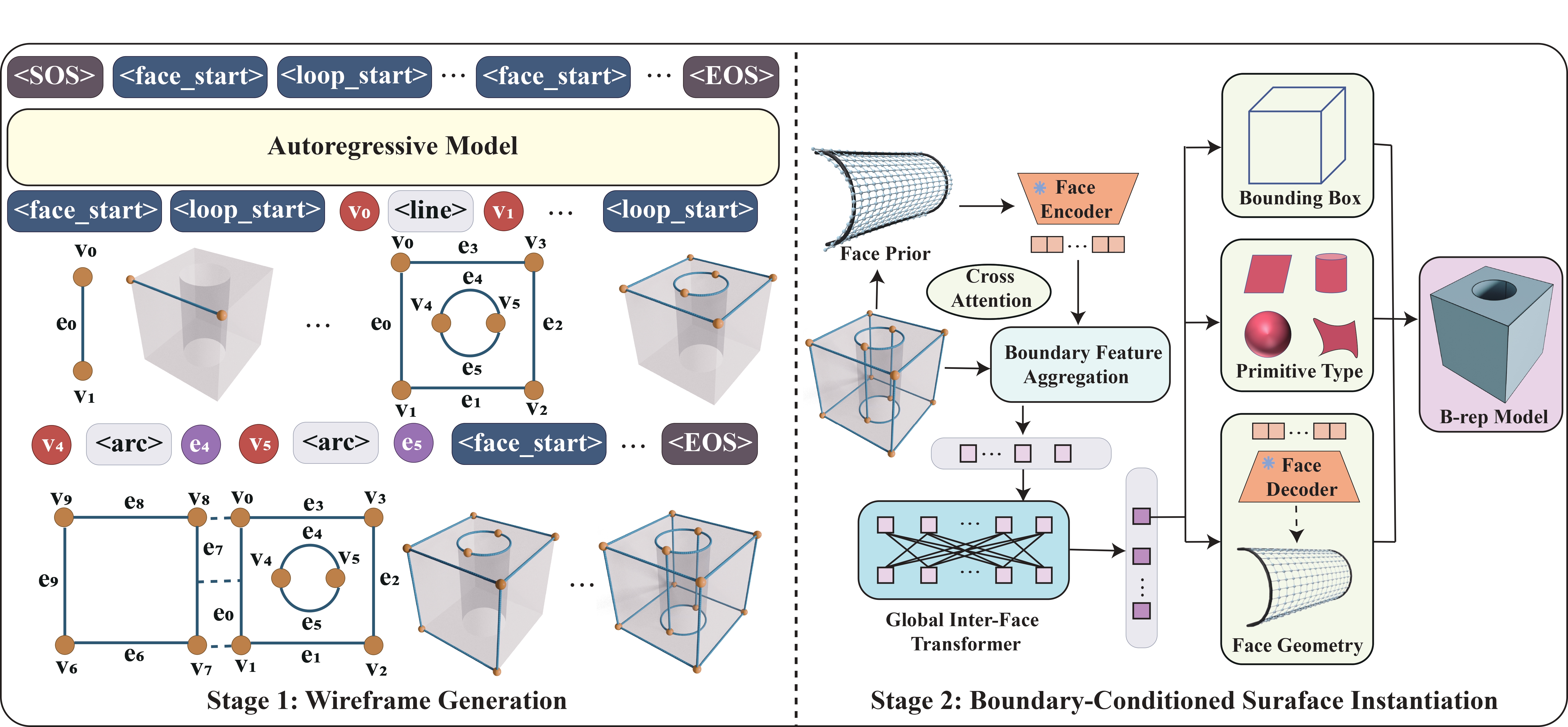}
  \caption{Pipeline of BrepForge. Our framework first autoregressively synthesizes a face-aware wireframe that explicitly encodes hierarchical Vertex–Edge–Face (V–E–F) connectivity to ensure topological closure. Subsequently, the boundary-conditioned surface instantiation module populates the scaffold with precise geometries by integrating analytical priors with global inter-face refinement.}
  \label{fig: pipeline}
  \vspace{-6pt}
\end{figure*}

\section{Autoregressive Face-Aware Wireframe Generation}
\label{sec: wireframe}

Fig.~\ref{fig: pipeline} illustrates the overall pipeline of BrepForge. In the first stage, we generate a topologically consistent wireframe via an autoregressive model. Unlike prior work that represents wireframes as unordered edge sets, we adopt a face-aware serialization scheme that organizes the structure into ordered sequences of boundary loops, encoded as interleaved vertex–edge tokens. This design explicitly captures the hierarchical Vertex–Edge–Face (V–E–F) relationships and ensures topological closure during generation. In the second stage, BrepForge instantiates surface geometries conditioned on the generated wireframe. Boundary features are fused with analytic geometric priors to construct initial face-level embeddings, which are further processed by a Transformer-based network to enable global information exchange and enforce consistency across faces. Finally, a lightweight MLP head predicts refined surface geometries from the resulting context-aware embeddings.

We next detail the wireframe generation process in this section, followed by the surface instantiation stage in Sec.~\ref{sec: faceGeom_gen}.

\subsection{Vertex and Edge Tokenization}
\label{sec: quant}
To enable autoregressive modeling, we discretize wireframe geometry into a finite token vocabulary. Each B-rep model is first normalized to a unit bounding box $[-1,1]^3$ to establish a consistent coordinate frame across the dataset.

Each vertex position $\mathbf{v}_i = (x_i, y_i, z_i) \in \mathbb{R}^3$ is quantized using a uniform $10$-bit grid along each axis, mapping continuous coordinates to integer indices in $\{0, 1, \dots, 1023\}$. Vertices that fall into the same grid cell are merged to preserve topological consistency. Samples with irrecoverable conflicts introduced by quantization are discarded to ensure data quality.

Representing edges with heterogeneous parametric types (e.g., lines, circular arcs, and B-splines) under a unified tokenization scheme is challenging. Prior approaches based on uniform discrete sampling \cite{jayaraman2021uv, xu2024brepgen} or spline fitting \cite{li2025dtgbrepgen} often suffer from over-parameterization for simple primitives or reduced fidelity for complex geometries.
To balance compactness and expressiveness, we adopt a hybrid geometric encoding strategy. For simple canonical primitives (lines and circular arcs), we avoid explicit geometric encoding at this stage and instead defer their parameterization to the adaptive serialization process (Sec.~\ref{sec: serialization}), where they can be compactly specified by boundary constraints. For complex geometries, we first normalize each curve $\mathbf{e}_i$ to the $[-1,1]^3$ domain and uniformly sample $N=32$ points along its parametric domain, forming a point set $\mathcal{P}_i \in \mathbb{R}^{N \times 3}$. This point set is then compressed into a sequence of 12 discrete tokens using a Residual Quantized VAE (RQ-VAE) \cite{lee2022autoregressive}, with each token indexing a shared codebook of size $s$.

\subsection{Adaptive Face-Aware Serialization}
\label{sec: serialization}

A fundamental challenge in autoregressive B-rep modeling—whether for full-shape synthesis or wireframe generation—is the linearization of the coupled geometry-topology structure into a 1D sequence. BrepGPT \cite{li2025brepgpt} collapses local topological hierarchy into vertex-level attributes, leading to an information-dense representation that introduces ambiguity during topological decoding. In contrast, AutoBrep \cite{xu2025autobrep} adopts a global BFS traversal, which results in non-intuitive structural ``jumps'' and long-range dependencies. As a consequence, these methods force the model to prioritize learning complex,
implicit traversal patterns rather than the underlying geometric
regularities of the CAD model.

To address this issue, we propose a face-aware serialization strategy that follows a more natural geometric organization. We temporarily relax global manifold constraints by treating shared edges as co-located entities, allowing the wireframe to be decomposed into a collection of independent faces. Each face $S_i$ is serialized as an ordered sequence of boundary loops, with a canonical orientation: the outer loop is traversed counter-clockwise, while inner loops (holes) are traversed clockwise. Within each loop, we perform an interleaved vertex–edge traversal starting from a designated pivot vertex $\mathbf{v}_1^i$. The resulting serialized sequence for face $S_i$ is defined as:
\begin{equation}
\label{eq: seq}
    S_i = [\langle \texttt{face\_start} \rangle, \langle \texttt{loop\_start} \rangle, \mathbf{v}_1^i, \langle \texttt{edge\_type} \rangle, \mathbf{e}_1^i, \mathbf{v}_2^i, \cdots],
\end{equation}
where $\langle \texttt{face\_start} \rangle$ and $\langle \texttt{loop\_start} \rangle$ serve as structural delimiters to define the hierarchical boundaries of faces and loops. Each vertex $\mathbf{v}_j^i$ is represented by its three quantized coordinate tokens in $x, y, z$ order. A central design choice of our serialization is the tiered edge representation, which introduces an edge type token $\langle \texttt{edge\_type} \rangle \in \{\langle \texttt{line} \rangle, \langle \texttt{arc} \rangle, \langle \texttt{complex} \rangle\}$ to adaptively balance expressiveness and compactness:
\begin{itemize}[leftmargin=2em]
    \item \textbf{Lines} $\langle \texttt{line} \rangle$: Given that a line segment is uniquely determined by its endpoints ($\mathbf{v}_j^i$ and $\mathbf{v}_{j+1}^i$), we dispense with additional geometric tokens ($\mathbf{e}_j^i = \emptyset$). This allows edge connectivity to be implicitly determined by the ordered vertex sequence.
    \item \textbf{Circular Arcs} $\langle \texttt{arc} \rangle$: An arc is parameterized by its 10-bit quantized midpoint ($\mathbf{e}_j^i \in \{0, \dots, 1023\}^3$), following the same quantization scheme as vertices. Combined with the endpoints ($\mathbf{v}_j^i$ and $\mathbf{v}_{j+1}^i$), this triplet of points ensures the unique reconstruction of the circular arc.
    \item \textbf{Complex Curves} $\langle \texttt{complex} \rangle$: For all other complex geometries (e.g., ellipses, parabolas, and B-splines), we utilize the $12$ discrete tokens generated by our RQ-VAE (Section~\ref{sec: quant}) to preserve high-fidelity geometric details ($\mathbf{e}_j^i \in \{0,1,\dots,s-1\}^{12}$).
\end{itemize}
This adaptive encoding leverages geometric priors of CAD models, where analytic lines and arcs are predominant. By avoiding the over-parameterization of simple segments, we substantially compress the sequence length. Fig.~\ref{fig: pipeline} illustrates this process within our overall pipeline, using a face with concentric loops as a representative serialization example.

To ensure deterministic serialization, we adopt a hierarchical sorting scheme following \cite{hao2024meshtron, li2025brepgpt}: (1) Global vertices are sorted lexicographically by $(z, y, x)$ coordinates to assign unique indices; (2) Faces are ordered lexicographically based on their sorted constituent vertex indices; (3) Within each loop, the vertex with the lowest global index is chosen as the pivot $\mathbf{v}_1^i$. Following this canonical order, individual face sequences are concatenated to form the complete wireframe representation:
\begin{equation}
\mathcal{S} = S_1 \oplus S_2 \oplus \dots \oplus S_{N_f},
\end{equation}
where $N_f$ is the total number of faces and $\oplus$ denotes sequence concatenation.

This face-aware serialization simplifies the generative task by decomposing complex global topologies into a series of local, repetitive geometric units, and encourages the model to produce consistent geometries for shared entities across adjacent faces. Consequently, the global V–E–F topology can be recovered through a simple merging process: (1) vertices with identical quantized coordinates are merged; and (2) edges whose merged endpoints coincide and whose geometries fall within a predefined tolerance are consolidated.

\subsection{Hierarchical Structural Embedding}
\label{sec: pe}
Due to the deeply nested structure of B-reps, identical geometric or topological entities may appear at vastly different absolute positions across different wireframes. To mitigate this issue, we introduce a hierarchical structural embedding that encodes each token by its role within the evolving B-rep hierarchy. Specifically, each token $\mathcal{S}_k$ in the wireframe sequence $\mathcal{S}$ is associated with a structural multi-index defined as:
\begin{equation}
    \mathcal{I}(\mathcal{S}_k | \mathcal{S}_{<k})=(\mathcal{I}_{\text{face}}, \mathcal{I}_{\text{loop}}, \mathcal{I}_{\text{type}}, \mathcal{I}_{\text{geom}}, \mathcal{I}_{\text{intra}}),
\end{equation}
where each component is deterministically derived from the hierarchical state induced by the prefix $\mathcal{S}_{<k}$, without accessing future tokens. The index components are defined as follows:
\begin{itemize}[leftmargin=1.5em]
\item \textbf{Face and Loop Indices} $\mathcal{I}_{\text{face}} \in \{0, 1, \dots, M_f\}$ together with $\mathcal{I}_{\text{loop}} \in \{0, 1, \dots, M_l\}$: These indices identify the parent face and its constituent loop, respectively, establishing the global topological context. Here, $M_f$ and $M_l$ denote the predefined maximum capacities for faces per model and loops per face. 
\item \textbf{Entity Category} $\mathcal{I}_{\text{type}} \in \{0, 1, 2\}$: A categorical variable indicating the entity type, where $1$ corresponds to a vertex, $2$ to an edge, and $0$ denotes a null entity.
\item \textbf{Geom-Entity Index} $\mathcal{I}_{\text{geom}} \in \{0, 1, \dots, N_g\}$: This index specifies the relative sequential position of a vertex or edge within its parent loop, where $N_g$ denotes the maximum allowable entities per loop.
\item \textbf{Intra-Entity Offset} $\mathcal{I}_{\text{intra}} \in \{0, 1, \dots, 12\}$: A fine-grained offset capturing the internal structure of an entity. For vertices, it distinguishes between the $(x, y, z)$ coordinates. For edges, it identifies the edge type or the specific index within the 12-token RQ-VAE geometry sequence.
\end{itemize}
We reserve a null index $\{0\}$ to indicate non-applicable hierarchical levels within the structural embedding. For high-level structural delimiters such as $\langle \texttt{loop\_start} \rangle$, only the relevant hierarchy components (i.e., $\mathcal{I}_{\text{face}}$ and $\mathcal{I}_{\text{loop}}$) are assigned valid indices, while all lower-level components are set to the null index. This hierarchical masking mechanism explicitly embeds topological depth into the token representation, allowing the model to distinguish high-level layout markers from low-level geometric attributes.

The final input embedding for each token $\mathcal{S}_k$ is obtained by element-wise summation of its semantic vocabulary embedding and the five learnable structural embeddings. By encoding relative structural roles instead of absolute positions, this design introduces structural invariance into the model, enabling it to recognize recurring geometric and topological patterns across different wireframes. 

\section{Boundary-Conditioned Surface Instantiation}
\label{sec: faceGeom_gen}
Conditioned on the synthesized wireframe scaffold, we reformulate face geometry generation as a structured refinement process. We first introduce the face geometry representation, followed by the construction of an initial face prior from boundary information, and finally describe a Transformer-based refinement network for predicting the geometry of each face.

\subsection{Symmetry-Invariant Face Representation}
\label{sec: faceGeom}
Consistent with our edge representation, we adopt a grid-based sampling strategy to encode diverse face geometries into a fixed-length latent space. For each face, we first normalize its geometry into the unit cube $[-1,1]^3$, and then uniformly sample an $N \times N$ grid ($N=32$) over the bounding box of its UV domain. This yields a discrete representation $\mathbf{f}_i \in \mathbb{R}^{N \times N \times 3}$ that captures surface geometry while remaining agnostic to specific parametric formulations.

Due to the inherent $D_4$ symmetry of the UV domain, a single physical surface can correspond to eight grid arrangements, leading to inconsistent latent representations. To eliminate this ambiguity, we introduce a deterministic geometric canonicalization that aligns each grid to a unique orientation prior to encoding. First, the grid is rotated and reflected such that the quadrant with the maximum $L_1$ energy (defined as the sum of absolute coordinate values) is aligned to the upper-left quadrant. For centrally symmetric geometries where the energy distribution is degenerate, we further resolve ambiguity by lexicographically ranking the quantized centroids of the four quadrants. This hybrid approach ensures a unique, noise-robust orientation even for highly symmetric primitives or noisy prior grids. Each face $\mathbf{f}_i$ is then compressed into a compact $48$-dimensional latent code $\mathbf{z}_i \in \mathbb{R}^{48}$ using a CNN-based VAE, ensuring that the refinement network (Sec.~\ref{sec: network}) receives consistent geometric signals regardless of the initial wireframe orientation.

\subsection{Analytical Face Prior Generation}
\label{sec: priors}

By deriving an analytic face prior from boundary information, we reformulate face generation as a structured refinement process. The generation workflow consists of three primary stages:

\textit{Robust Local Basis Construction}. Given the boundary points, we first estimate a stable surface normal using Newell's Algorithm. This method computes the normal by aggregating the signed areas of the loop's projections onto the Cartesian planes, providing superior robustness against noise or irregular point distributions. We then establish an orthogonal local basis by projecting the boundary points onto the tangent plane and aligning the axes with the principal components of the projected distribution.

\textit{Surface Approximation}. We fit a quadratic surface to the boundary points in the local basis, providing a flexible yet constrained low-order model that captures the face's underlying curvature.

\textit{Grid Sampling}. Finally, we identify the rectangular bounding box of the outer loop in the local planar system and uniformly sample an $N \times N$ ($N=32$) grid. By evaluating the fitted quadratic function at each sample site and normalizing the coordinates to the $[-1, 1]^3$ cube, we obtain the final prior $\mathbf{f}_i^{\text{prior}} \in \mathbb{R}^{N\times N\times 3}$. Notably, we sample the entire rectangular domain irrespective of interior voids (inner loops). The precise topological trimming is implicitly preserved by the boundary loops, which is ultimately resolved by the CAD kernel during the final reconstruction.


\subsection{Face Geometry Refinement Network}
\label{sec: network}
With the analytical face priors and boundary loops established, we introduce a Transformer-based architecture to refine the initial estimates into high-fidelity B-rep geometries.

\textit{Local Boundary Encoding}. To capture geometric constraints of a wireframe, each boundary edge of face $\mathbf{f}_i$ is encoded using a dual representation. A 1D convolution processes the normalized sample points $\mathbf{e}_j^i$ to capture the local curvature profile, while an MLP encodes the global vertex pair $(\mathbf{v}_j^i, \mathbf{v}_{j+1}^i)$ to provide absolute spatial coordinates. This dual path ensures the network extracts both local geometric features and their relative placement within the assembly. These features are concatenated into a comprehensive edge embedding, enabling the network to perceive boundary shapes within their global context.

\textit{Prior-Guided Feature Aggregation}. To condense the variable number of boundary edge features into a fixed-dimensional face descriptor, we employ a Cross-Attention mechanism. The analytical prior $\mathbf{f}_i^{\text{prior}}$ is first encoded via the frozen VAE encoder into a latent code $\mathbf{z}_i^{\text{prior}}$, which serves as the Query $\mathbf{q}_i$. The previously encoded boundary edge features act as the Key $\mathbf{K}_i$ and Value $\mathbf{V}_i$, allowing the network to attend to relevant geometric constraints from the boundaries based on the initial surface estimate. To balance contributions from the prior and boundary features, we adopt a Gated Residual Connection \cite{dauphin2017language}:
\begin{equation}
\mathbf{h}_i = \text{LayerNorm}(\mathbf{q}_i + g \cdot \text{Attn}(\mathbf{q}_i, \mathbf{K}_i, \mathbf{V}_i)),
\end{equation}
where the gate $g$ regulates information flow according to the compatibility between the prior and boundary details. This design enables topologically grounded yet flexible refinement of the initial surface estimate.

\textit{Global Inter-Face Transformer}. To achieve global geometric coherence across the entire B-rep model, we treat the face descriptors $\{\mathbf{h}_i\}$ as a sequence of tokens and process them through a Self-Attention mechanism. Additionally, a prior bounding box is derived for each face from its boundary loops and used as a spatial positional encoding added to the face tokens, enabling the Transformer to reason about the relative 3D layout of the assembly. After global reasoning, the refined face tokens are processed by three specialized prediction heads: (1) a Geometry Head that regresses the latent code $\mathbf{z}_i$ for VAE decoding; (2) a Primitive Head that classifies each face as \textit{Plane}, \textit{Cylinder}, \textit{Sphere}, or \textit{Complex}; and (3) a Bounding Box Head that predicts the refined bounding boxes. The model is trained using a multi-task loss:
\begin{equation}
\mathcal{L} = \mathcal{L}_{\text{geom}} + \lambda_{\text{box}}\mathcal{L}_{\text{box}} + \lambda_{\text{type}}\mathcal{L}_{\text{type}},
\label{eq: loss}
\end{equation}
where $\mathcal{L}_{\text{geom}}$ is the Mean Squared Error (MSE) for latent code regression, $\mathcal{L}_{\text{box}}$ is the Smooth L1 loss for bounding box refinement, and $\mathcal{L}_{\text{type}}$ is the Cross-Entropy loss for primitive classification. The weights $\lambda_{\text{box}}$ and $\lambda_{\text{type}}$ balance the relative contributions of individual objectives.

During inference, synthesized face grids are first transformed from their local canonical frames into the global coordinate system using the predicted bounding boxes. Type-specific surface fitting is then performed: standard primitives (\textit{Plane}, \textit{Cylinder}, \textit{Sphere}) are reconstructed using analytic models, while \textit{Complex} faces are fitted with B-spline surfaces. Finally, the fitted surfaces are integrated with the wireframe predicted in the first stage and assembled into a solid model using the OpenCascade kernel \cite{pyOCCT}.

\section{Experiments}
\subsection{Implementation Details}
\textit{Dataset}. We evaluate our method on the ABC dataset \cite{koch2019abc}. Our experiments focus on single-body models, and we exclude extreme cases with serialized sequences exceeding 8,000 tokens. To avoid geometric ambiguities when constructing analytical face priors from boundary loops, all periodic surfaces are split along their seam edges. Following \cite{xu2024brepgen}, we apply a strict deduplication protocol to remove near-duplicate shapes. After filtering, the resulting dataset contains approximately 280k models, which are randomly divided into training, validation, and test sets with a 90/5/5\% split.

\textit{Network Architecture}. For edge geometry compression, we employ a 1D CNN-based RQ-VAE with $L=3$ quantization layers. All layers share a codebook of size $s=256$, where each code has a dimensionality of 64. For face geometry encoding, we adopt a KL-regularized convolutional VAE that maps sampled face grids to a compact 48-dimensional latent representation. The wireframe generation module is implemented as a decoder-only Transformer with 16 layers, a hidden dimension of 1024, and 16 attention heads. Each Transformer layer consists of a self-attention block and is optionally augmented with a cross-attention module to incorporate conditional inputs. For global inter-face reasoning, we adopt a Transformer architecture that mirrors the wireframe generation module, but with a lightweight configuration of 6 layers.

\textit{Training}. For the RQ-VAE, the codebook is initialized via k-means, and the model is trained for approximately 20 hours on a single NVIDIA A800 GPU with a learning rate of $5 \times 10^{-5}$ and a quantization loss weight of $0.25$. The face VAE is trained on the same hardware for 24 hours using a learning rate of $5 \times 10^{-4}$ and a KL-divergence weight of $10^{-6}$. For both the wireframe generation and global inter-face Transformers, we use a learning rate of $10^{-4}$ with a 1,000-step linear warm-up starting from $10^{-8}$, along with a weight decay of $10^{-5}$. These models are trained on a cluster of eight NVIDIA A800 GPUs for 4 days to ensure full convergence. Other hyper-parameters are set to $M_f=70, M_l=15, N_g=30, \lambda_{box}=0.4,$ and $\lambda_{type}=0.2$.

\textit{Evaluation Metrics}. Following \cite{xu2024brepgen, li2025dtgbrepgen}, we evaluate generation quality using two categories of metrics. (1) Distribution Metrics: Coverage (COV), Minimum Matching Distance (MMD), and Jensen–Shannon Divergence (JSD). COV and MMD measure the completeness and accuracy of the generated set with respect to the ground truth, while JSD quantifies the similarity between their overall spatial distributions. (2) CAD Metrics: Novel, Unique, and Valid. \textit{Novel} evaluates the model’s ability to generalize beyond the training data, and \textit{Unique} measures the diversity within the generated set. \textit{Valid} reports the proportion of watertight solids, as verified using the \texttt{BRepCheck\_Analyzer().IsValid()} function from the OpenCASCADE kernel \cite{pyOCCT}. We further report Cyclomatic Complexity (CC) \cite{contero2023quantitative} to measure the structural complexity of the generated B-rep models.

\subsection{Wireframe Generation}
\begin{table}[tb]
\centering
\captionsetup{skip=4pt}
\caption{Quantitative comparison of unconditional wireframe generation. Both MMD and JSD scores are multiplied by $10^2$.}
\label{tab: wire_uncond}
\begin{tabular}{lcccccc}
\toprule
\multirow{2}{*}{Method} 
& \multicolumn{2}{c}{COV (\%$\uparrow$)} 
& \multicolumn{2}{c}{MMD ($\downarrow$)} 
& \multicolumn{2}{c}{JSD ($\downarrow$)} \\
\cmidrule(lr){2-3}
\cmidrule(lr){4-5}
\cmidrule(lr){6-7}
& CD & EMD & CD & EMD & CD & EMD \\
\midrule
BrepGen
& 66.38 & 66.91 & 3.36 & 4.80 & 3.54 & 5.11 \\
DTGBrepGen
& 65.72 & 67.05 & 3.24 & 4.88 & 3.57 & 5.03 \\
CLR\_Wire
& 72.17 & 73.45 & 3.01 & 4.72 & 3.26 & 4.93 \\
\textbf{Ours} 
& \textbf{76.27} & \textbf{77.26} & \textbf{2.65} & \textbf{4.38} & \textbf{2.90} & \textbf{4.61} \\
\bottomrule
\end{tabular}
\end{table}

\begin{table}[tb]
\centering
\captionsetup{skip=4pt}
\caption{Quantitative evaluation of point cloud-conditioned wireframe reconstruction. Both CD and EMD scores are multiplied by $10^2$.}
\label{tab: wire_cond}
\begin{tabular}{lccc}
\toprule
Method & CD ($\downarrow$) & EMD ($\downarrow$) & F-score ($\uparrow$) \\
\midrule
RFEPS 
& 1.27 & 8.41 & 0.88 \\
NerVE
& 1.23 & 9.62 & 0.91 \\
\textbf{Ours} 
& \textbf{0.76} & \textbf{2.11} & \textbf{0.93} \\
\bottomrule
\end{tabular}
\end{table}

\begin{figure}[tb]
  \centering
  \includegraphics[width=0.72\linewidth]{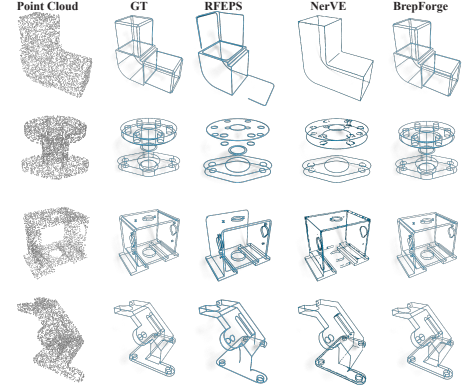}
  \captionsetup{skip=4pt}
  \caption{Qualitative comparison of point-cloud conditioned wireframe reconstruction. BrepForge produces topologically complete wireframes that align precisely with the sparse input (4,096 points), whereas baselines often struggle with structural connectivity.}
  \label{fig: wire_pc}
    \vspace{-2pt}
\end{figure}

\textit{Unconditional Generation.} We first evaluate the unconditional autoregressive wireframe generation performance of BrepForge against BrepGen \cite{xu2024brepgen}, DTGBrepGen \cite{li2025dtgbrepgen}, and CLR-Wire \cite{ma2025clr} using Distribution-based metrics. We conduct 10 independent evaluation trials. In each trial, we randomly sample 3,000 generated wireframes and 1,000 ground-truth wireframes, and report metrics averaged across all trials. All metrics are computed using both Chamfer Distance (CD) and Earth Mover’s Distance (EMD). As shown in Tab.~\ref{tab: wire_uncond}, our method achieves higher COV and lower MMD scores, indicating improved diversity and geometric fidelity of the generated wireframe population. Moreover, the consistently lower JSD values suggest a closer alignment between the output distribution of our model and the underlying CAD shape manifold.

\textit{Point Cloud-Conditioned Reconstruction.} The flexible architecture of BrepForge supports conditional synthesis, which we evaluate through point cloud-conditioned reconstruction experiments. For each input geometry, we uniformly sample 4,096 points as the conditional constraint. These points are encoded into 256 latent tokens via a frozen point encoder \cite{zhao2023michelangelo} and injected into the Transformer through cross-attention. We compare our approach with RFEPS \cite{xu2022rfeps} and NerVE \cite{zhu2023nerve}. As illustrated in Fig.~\ref{fig: wire_pc}, BrepForge reconstructs topologically complete wireframes that align closely with the input point clouds, whereas baseline methods often produce fragmented segments or geometric distortions. Quantitative results in Tab.~\ref{tab: wire_cond}, including CD, EMD, and F-score, further demonstrate superior reconstruction precision and structural coherence. Notably, BrepForge achieves these results using only 4,096 points, significantly outperforming baseline methods that require dense inputs of 20,000 points.

\begin{table}[tb]
\centering
\captionsetup{skip=4pt}
\caption{Quantitative comparison of unconditional B-rep generation. Both MMD and JSD scores are multiplied by $10^2$. Ours$_{\text{w/o prior}}$ and Ours$_{\text{w/o refine}}$ denote variants without analytic geometric priors and surface refinement, respectively.}
\label{tab: brep-uncond}
\setlength{\tabcolsep}{3pt}
\begin{tabular}{lccc|ccc|c}
\toprule
Method 
& COV & MMD & JSD & Novel & Unique & Valid & CC \\
& (\%$\uparrow$) & ($\downarrow$) & ($\downarrow$)
& (\%$\uparrow$) & (\%$\uparrow$) & (\%$\uparrow$) & ($\uparrow$) \\
\midrule
BrepGen
& 66.17 & 1.65 & 1.74 & 99.6 & 96.6 & 60.0 & 28.9 \\
DTGBrepGen
& 65.99 & 1.63 & 1.61 & 99.7 & 96.4 & 64.6 & 32.6 \\
BrepDiff
& 66.03 & 1.69 & 1.58 & 99.7 & 96.0 & 62.5 & 32.1\\
Ours$_{\text{w/o refine}}$
& 59.45 & 2.01 & 1.97 & 99.1 & 95.6 & 57.4 & 35.3 \\
Ours$_{\text{w/o prior}}$
& 69.98 & 1.43 & 1.19 & \textbf{99.8} & 97.1 & 66.2 & 44.7 \\
\textbf{Ours}
& \textbf{72.65} & \textbf{1.24} & \textbf{0.92} & \textbf{99.8} & \textbf{97.3} & \textbf{70.9} & \textbf{47.2} \\
\bottomrule
\end{tabular}
\end{table}

\begin{figure*}[tb]
     \centering
     \begin{subfigure}[b]{0.3\textwidth}
         \centering
         \includegraphics[width=0.92\textwidth]{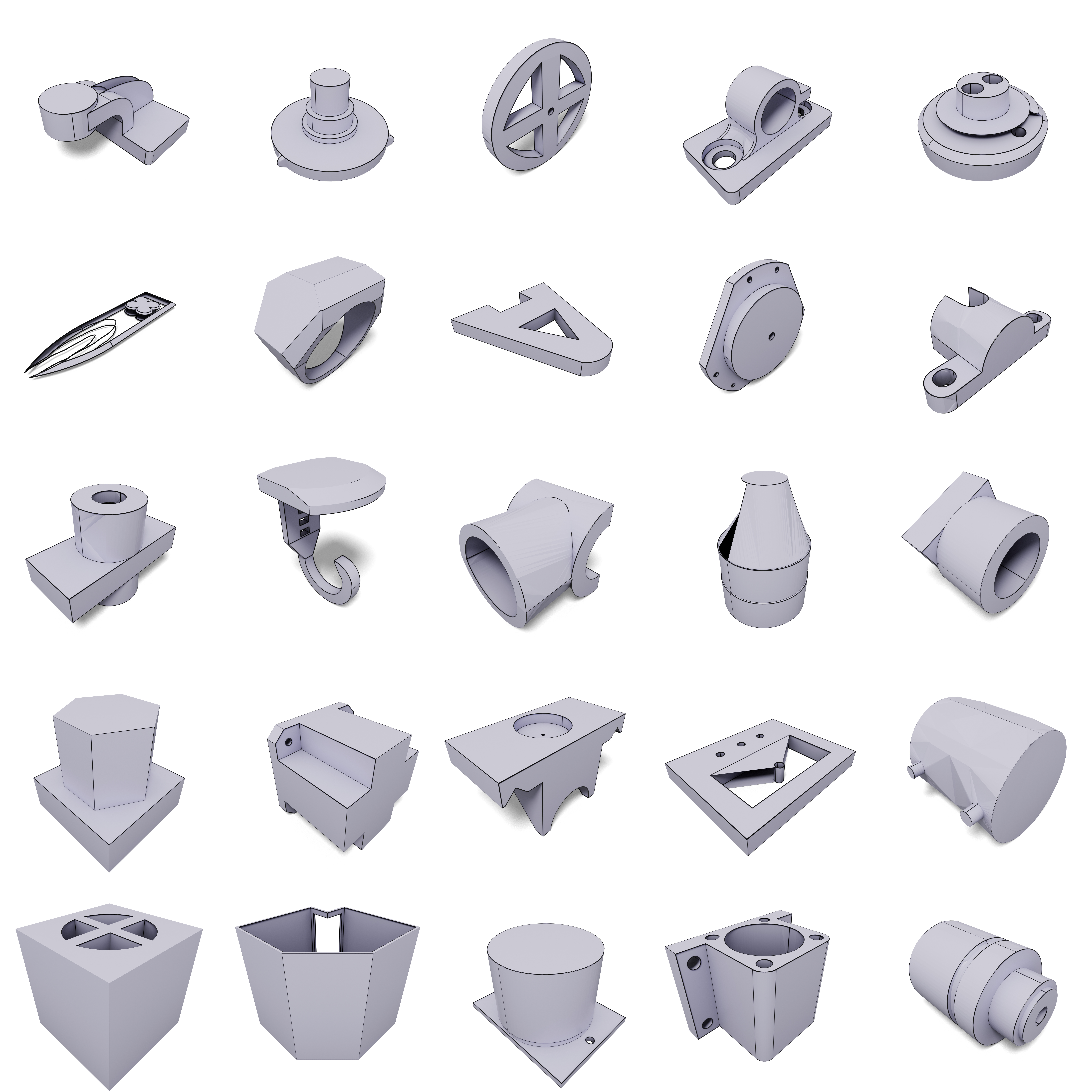}
         \caption{BrepGen}
         \label{fig:brepgen}
     \end{subfigure}
     \vrule
     \begin{subfigure}[b]{0.3\textwidth}
         \centering
         \includegraphics[width=0.92\textwidth]{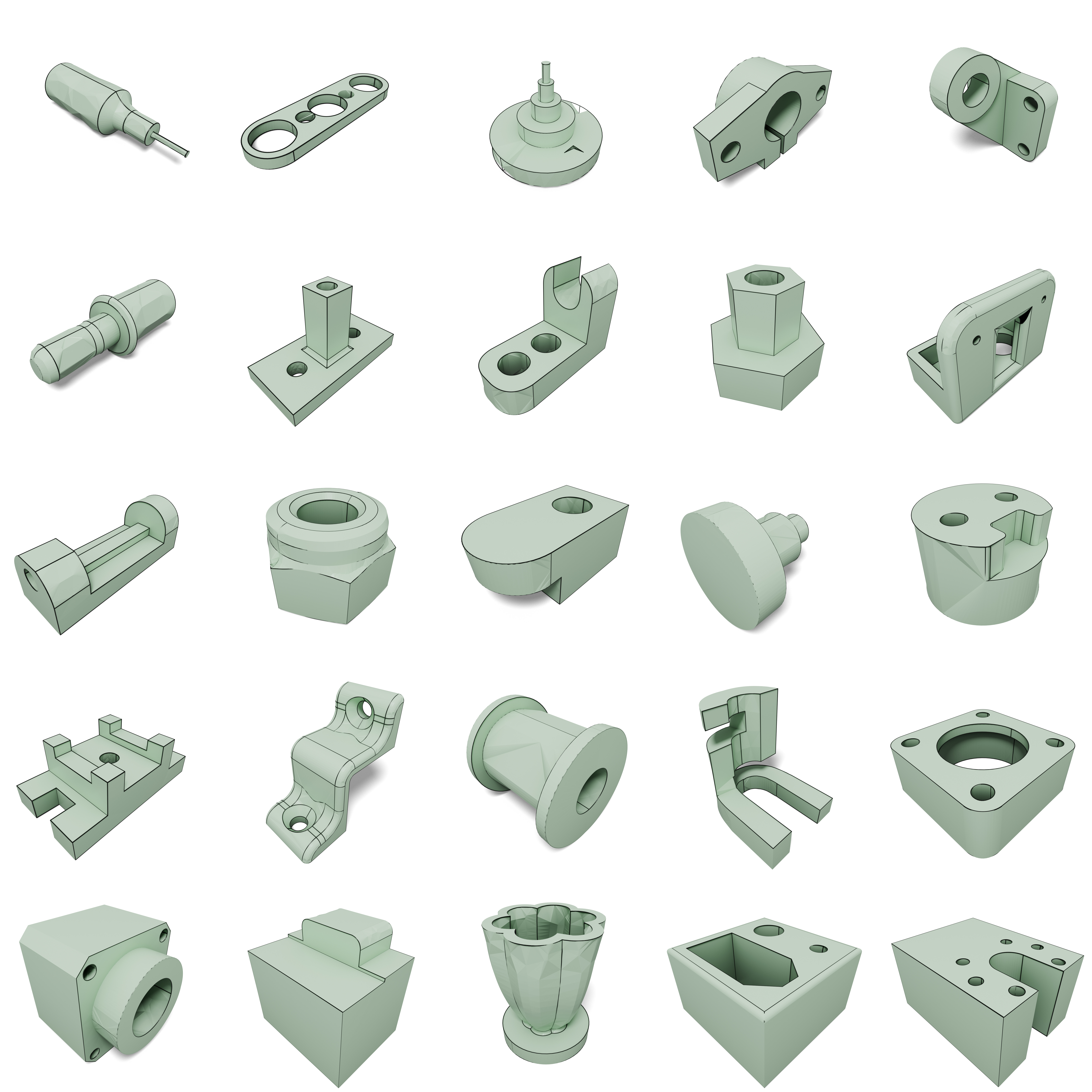}
         \caption{DTGBrepGen}
         \label{fig:dtg}
     \end{subfigure}
     \vrule
     \begin{subfigure}[b]{0.3\textwidth}
         \centering
         \includegraphics[width=0.92\textwidth]{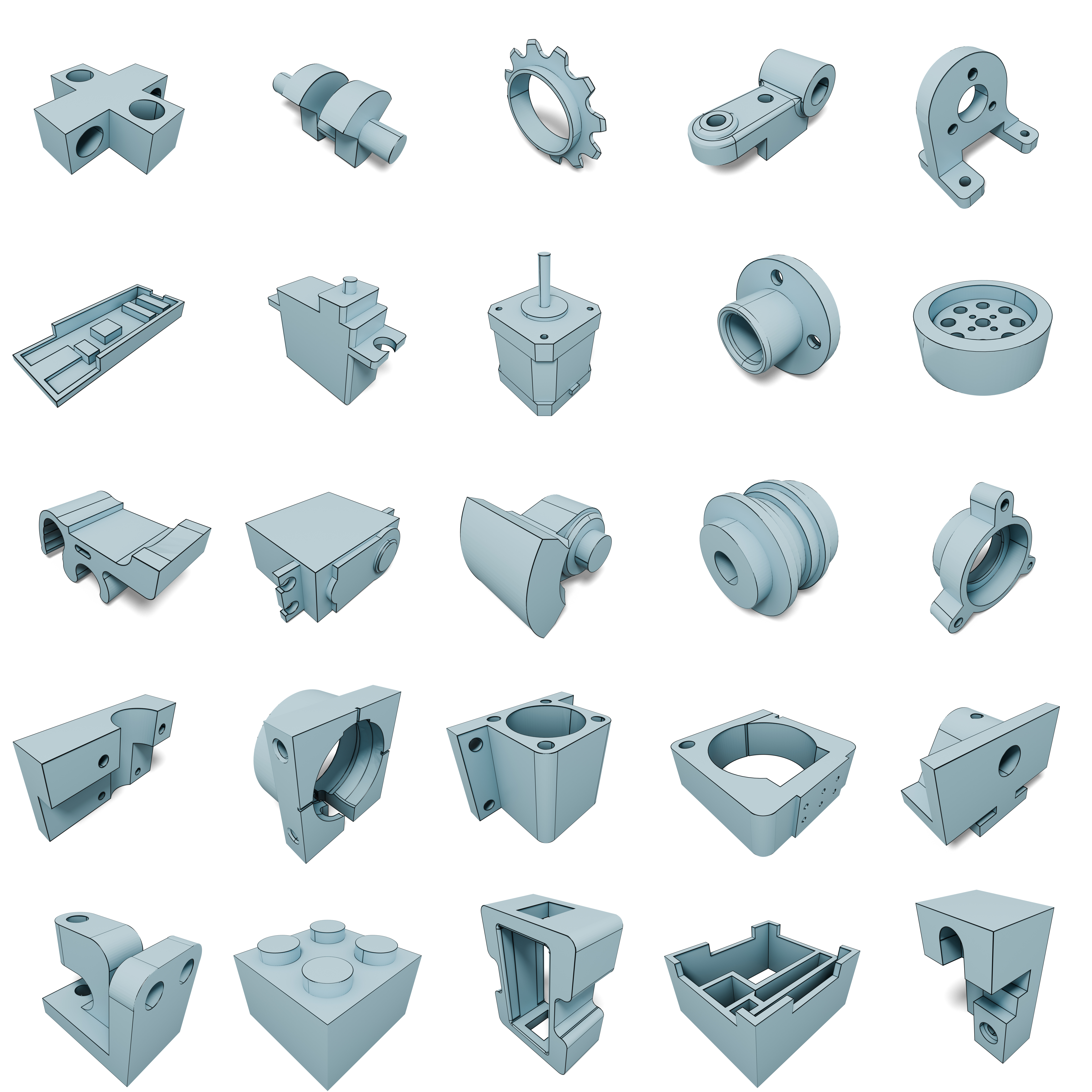}
         \caption{BrepForge}
         \label{fig:ours}
     \end{subfigure}
    \captionsetup{skip=4pt}
     \caption{Qualitative comparison of unconditional B-rep generation. Compared to existing baselines, BrepForge synthesizes models with significantly higher geometric complexity while consistently maintaining topological validity.}
     \label{fig: brep_uncond}
     \vspace{-6pt}
\end{figure*}

\subsection{B-rep Generation}

\textit{Unconditional Generation.} Given the limited availability of open-source B-rep generative frameworks, we benchmark BrepForge against BrepGen \cite{xu2024brepgen}, DTGBrepGen \cite{li2025dtgbrepgen}, and BrepDiff \cite{lee2025brepdiff}, all retrained on our dataset for fair comparison. As shown in Fig.~\ref{fig: brep_uncond}, our method excels at synthesizing models with richer geometric complexity and more intricate topological details, whereas baselines often produce over-simplified or fragmented surfaces. For quantitative evaluation, we follow the same protocol as in wireframe generation, conducting 10 trials with 3,000 generated samples and 1,000 ground-truth models each. As reported in Tab.~\ref{tab: brep-uncond}, BrepForge consistently outperforms all baselines across metrics, indicating a better fit to the underlying B-rep data manifold and enabling the synthesis of plausible and diverse CAD models. Notably, the high validity rate confirms that our face-aware serialization promotes consistent predictions for shared entities, facilitating coherent topology recovery through simple merging. Furthermore, elevated Cyclomatic Complexity (CC) scores indicate that BrepForge generates models with richer topological structures, reflecting its ability to capture complex B-rep configurations.

\begin{figure*}[t]
     \centering
      \includegraphics[width=0.9\textwidth]{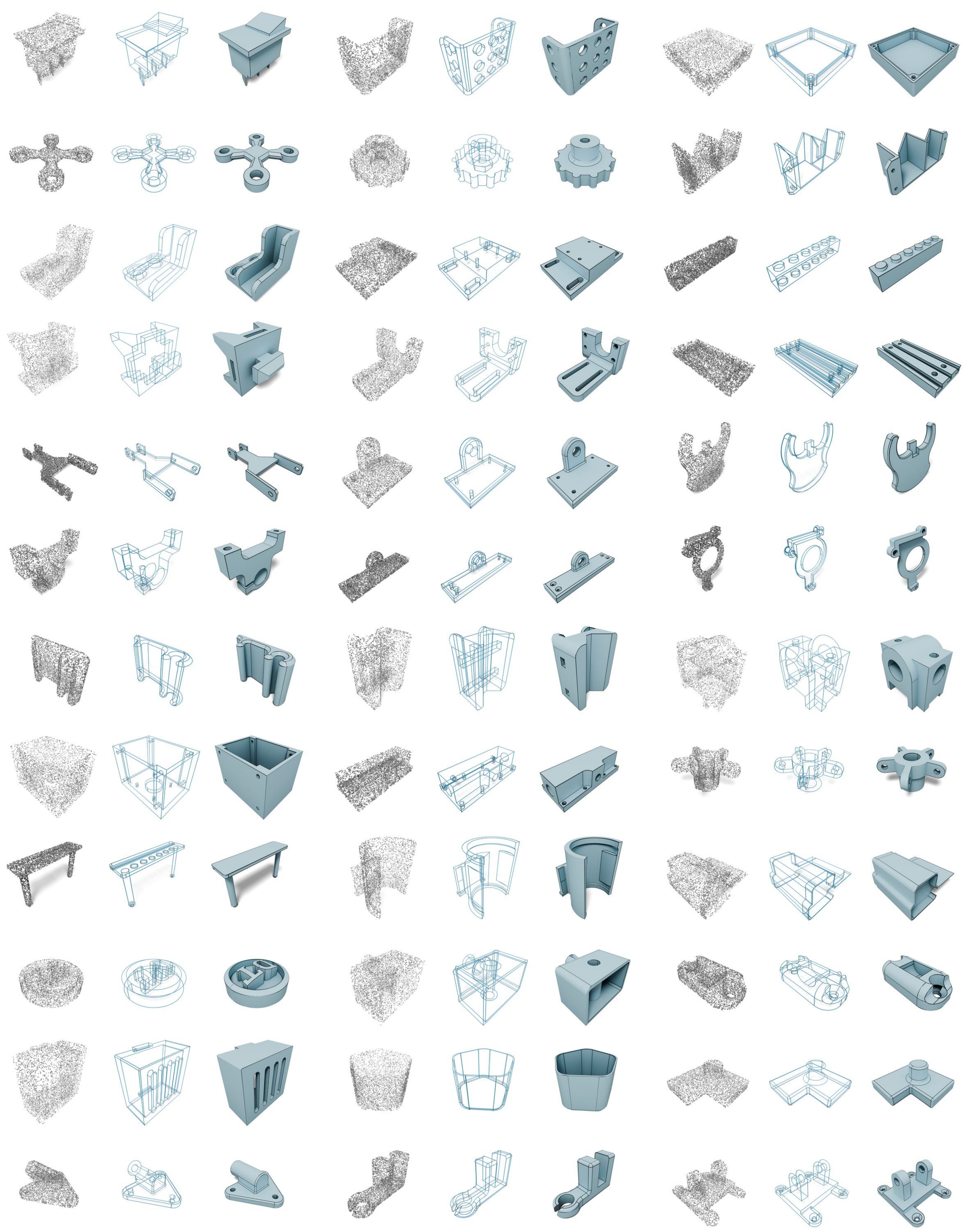}
     \caption{Gallery of point-cloud conditioned B-rep generation. For each example, we present the input point cloud, the synthesized wireframe scaffold, and the final instantiated B-rep model. Our method effectively recovers precise CAD geometries and valid topologies from sparse, discrete inputs.}
     \label{fig: pc}
\end{figure*}

\textit{Point Cloud-Conditioned Reconstruction.} Extending beyond wireframe reconstruction, we evaluate BrepForge’s capability for full B-rep synthesis under point cloud conditioning. As illustrated in Fig.~\ref{fig: pc}, our model synthesizes high-fidelity CAD models where generated surfaces align faithfully with both the wireframe scaffold and the original point samples. Even for objects with complex face transitions, BrepForge produces watertight solids with sharp features and precise curvature fitting. These results highlight the robustness of our architecture in transforming unstructured point samples into structured, topologically consistent B-reps.

\subsection{Analysis on Geometry Prior}
\label{sec: analysis}

\begin{figure}[tb]
  \centering
  \includegraphics[width=0.95\linewidth]{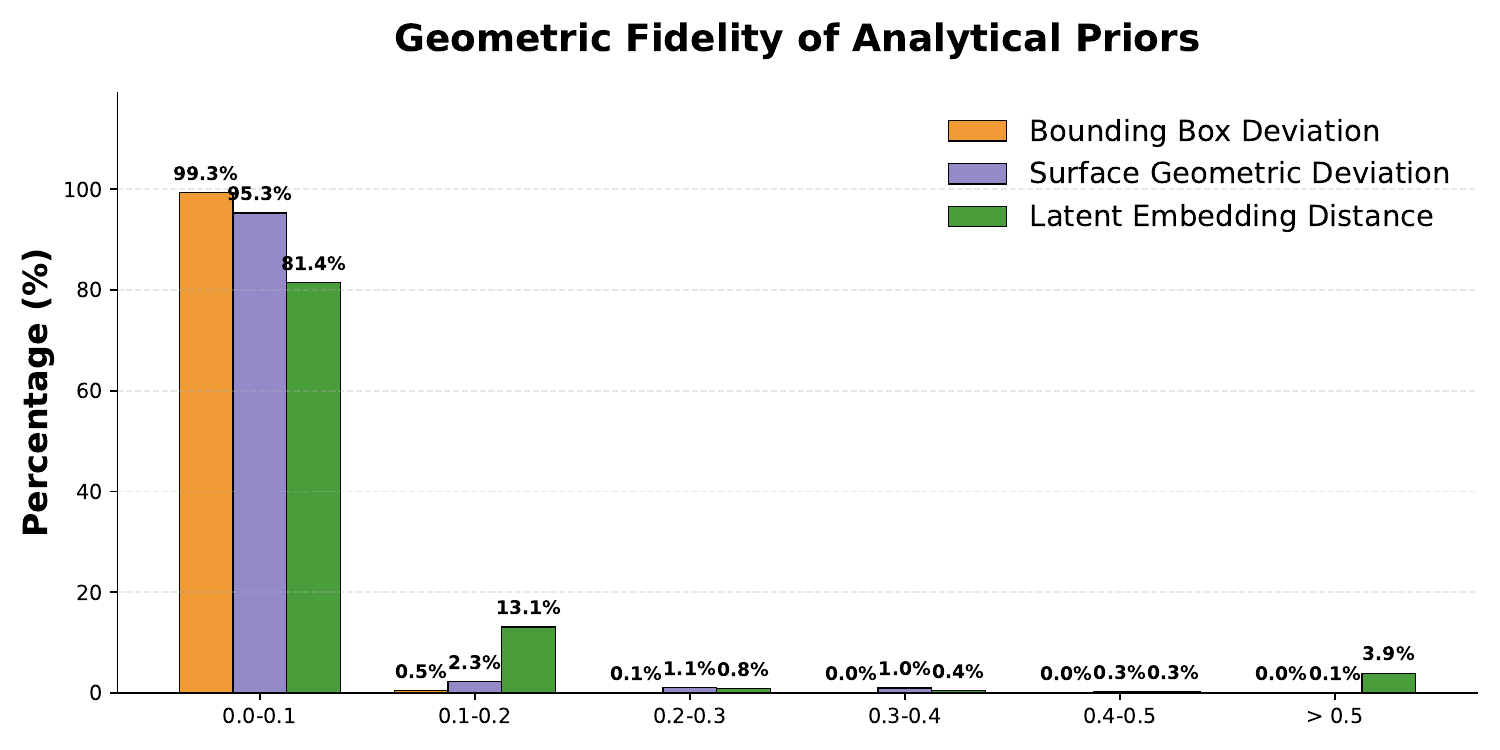}
  \captionsetup{skip=4pt}
  \caption{Geometric fidelity of analytical priors. We report the error distribution between the generated priors and ground-truth surfaces across three metrics: Bounding Box Deviation ($L_2$), Surface Geometric Deviation (CD), and Latent Embedding Distance ($L_2$).}
  \label{fig: distribution}
\end{figure}

\begin{figure*}[h]
    \centering
    \includegraphics[width=0.82\textwidth]{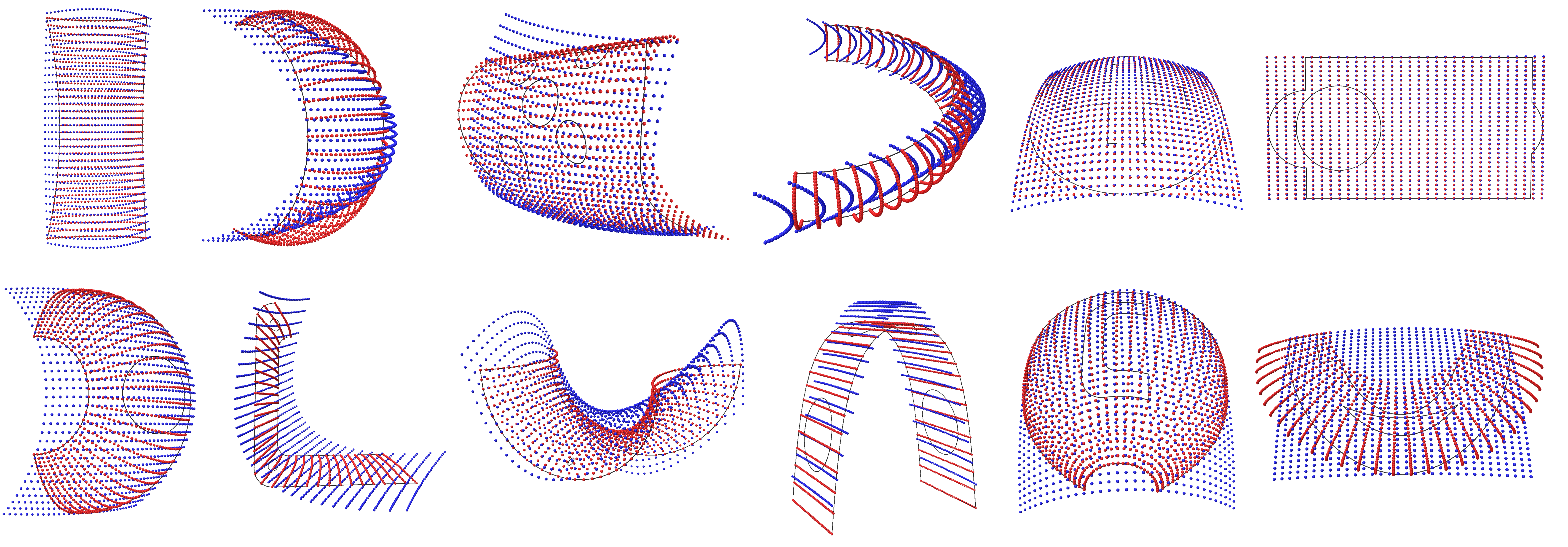}
    \caption{Qualitative evaluation of analytical priors. Our method produces initial surface grids (blue) that align closely with the boundary wireframes (black) and ground-truth geometry (red). The analytical priors effectively capture the underlying manifold curvature across diverse CAD surface types.}
    \label{fig: prior}
\end{figure*}

\textit{Quality of Analytical Priors.} 
We evaluate the fidelity of our analytical priors through visual and geometric comparisons. As shown in Fig.~\ref{fig: prior}, our method produces initial surface grids (blue) that are consistent with boundary wireframes (black) and closely approximate the ground-truth geometry (red). The priors capture the underlying curvature across diverse surface types, including cylindrical and high-curvature patches. Quantitatively, Fig.~\ref{fig: distribution} presents the error distribution between priors and ground-truth surfaces. The results demonstrate high precision: 99.3\% of bounding boxes and 95.3\% of surface CD errors fall within the $[0.0,0.1]$ range, with 81.4\% of latent embeddings also residing in this lowest error tier. This high-quality ``warm-start'' effectively constrains the geometric search space, allowing the subsequent Transformer to focus on fine-grained residuals rather than coarse global structure.

\textit{Ablation Study.} We conduct an ablation study on both the analytical prior and the surface refinement stage. Removing the prior input (denoted as Ours$_{\text{w/o prior}}$ in Tab.~\ref{tab: brep-uncond}) leads to a noticeable degradation in reconstruction accuracy. While the base architecture can regress surface geometry directly from wireframe tokens, incorporating analytic geometric priors provides a significant improvement in precision. We further evaluate a variant without the second-stage refinement (denoted as Ours$_{\text{w/o refine}}$), where the analytical priors are directly used as the final surface geometry. This variant results in substantially worse performance, highlighting the necessity of the refinement network for capturing fine-grained geometric details. 

\section{Conclusion and Future Work}
In this work, we presented BrepForge, a factorized framework that alleviates the representation bottleneck of monolithic B-rep synthesis. By exploiting the information asymmetry between wireframes and faces, BrepForge adopts a face-aware sequential encoding to generate topologically closed scaffolds and a structured refinement process to instantiate accurate surface geometries. Experimental results demonstrate that this two-stage factorization ensures both topological coherence and geometric fidelity in complex CAD modeling. Future work will focus on improving computational efficiency via more compact geometric representations to reduce autoregressive sequence length. We also plan to develop $D_4$-invariant face autoencoders to explicitly model geometric symmetries, eliminating orientation ambiguities during surface instantiation.


\begin{acks}
This work was supported by the Major Program of the National Natural Science Foundation of China under Grant No. 12494550 and Grant No. 12494555.
\end{acks}

\bibliographystyle{ACM-Reference-Format}
\bibliography{sample-bibliography}

@String{Computer = "{IEEE} Computer" }

@String{Chelsea = "Chelsea" }

@String{Springer = "Springer-Verlag" }

@book{weiler1986topological,
  title={Topological structures for geometric modeling (Boundary representation, manifold, radial edge structure)},
  author={Weiler, Kevin J},
  year={1986},
  publisher={Rensselaer Polytechnic Institute}
}

@inproceedings{wu2021deepcad,
  title={Deepcad: A deep generative network for computer-aided design models},
  author={Wu, Rundi and Xiao, Chang and Zheng, Changxi},
  booktitle={Proceedings of the IEEE/CVF International Conference on Computer Vision},
  pages={6772--6782},
  year={2021}
}

@article{xu2022skexgen,
  title={Skexgen: Autoregressive generation of cad construction sequences with disentangled codebooks},
  author={Xu, Xiang and Willis, Karl DD and Lambourne, Joseph G and Cheng, Chin-Yi and Jayaraman, Pradeep Kumar and Furukawa, Yasutaka},
  journal={arXiv preprint arXiv:2207.04632},
  year={2022}
}

@article{xu2023hierarchical,
  title={Hierarchical neural coding for controllable cad model generation},
  author={Xu, Xiang and Jayaraman, Pradeep Kumar and Lambourne, Joseph G and Willis, Karl DD and Furukawa, Yasutaka},
  journal={arXiv preprint arXiv:2307.00149},
  year={2023}
}

@article{xu2024cad,
  title={Cad-mllm: Unifying multimodality-conditioned cad generation with mllm},
  author={Xu, Jingwei and Wang, Chenyu and Zhao, Zibo and Liu, Wen and Ma, Yi and Gao, Shenghua},
  journal={arXiv preprint arXiv:2411.04954},
  year={2024}
}

@article{guo2025cadtrans,
  title={CADTrans: A code tree-guided CAD generative transformer model with regularized discrete codebooks},
  author={Guo, Xufei and Dong, Xiao and Cao, Juan and Chen, Zhonggui},
  journal={Graphical Models},
  volume={139},
  pages={101262},
  year={2025},
  publisher={Elsevier}
}

@article{khan2024text2cad,
  title={Text2cad: Generating sequential cad designs from beginner-to-expert level text prompts},
  author={Khan, Mohammad S and Sinha, Sankalp and Sheikh, Talha U and Stricker, Didier and Ali, Sk A and Afzal, Muhammad Z},
  journal={Advances in Neural Information Processing Systems},
  volume={37},
  pages={7552--7579},
  year={2024}
}

@article{willis2021fusion,
  title={Fusion 360 gallery: A dataset and environment for programmatic cad construction from human design sequences},
  author={Willis, Karl DD and Pu, Yewen and Luo, Jieliang and Chu, Hang and Du, Tao and Lambourne, Joseph G and Solar-Lezama, Armando and Matusik, Wojciech},
  journal={ACM Transactions on Graphics (TOG)},
  volume={40},
  number={4},
  pages={1--24},
  year={2021},
  publisher={ACM New York, NY, USA}
}

@article{jayaraman2022solidgen,
  title={Solidgen: An autoregressive model for direct b-rep synthesis},
  author={Jayaraman, Pradeep Kumar and Lambourne, Joseph G and Desai, Nishkrit and Willis, Karl DD and Sanghi, Aditya and Morris, Nigel JW},
  journal={arXiv preprint arXiv:2203.13944},
  year={2022}
}

@article{xu2024brepgen,
  title={Brepgen: A b-rep generative diffusion model with structured latent geometry},
  author={Xu, Xiang and Lambourne, Joseph and Jayaraman, Pradeep and Wang, Zhengqing and Willis, Karl and Furukawa, Yasutaka},
  journal={ACM Transactions on Graphics (TOG)},
  volume={43},
  number={4},
  pages={1--14},
  year={2024},
  publisher={ACM New York, NY, USA}
}

@inproceedings{li2025dtgbrepgen,
  title={DTGBrepGen: A Novel B-rep Generative Model through Decoupling Topology and Geometry},
  author={Li, Jing and Fu, Yihang and Chen, Falai},
  booktitle={Proceedings of the Computer Vision and Pattern Recognition Conference},
  pages={21438--21447},
  year={2025}
}

@inproceedings{li2025stitch,
  title={Stitch-A-Shape: Bottom-up Learning for B-Rep Generation},
  author={Li, Pu and Zhang, Wenhao and Chen, Jinglu and Yan, Dongming},
  booktitle={Proceedings of the Special Interest Group on Computer Graphics and Interactive Techniques Conference Conference Papers},
  pages={1--12},
  year={2025}
}

@inproceedings{lee2025brepdiff,
  title={Brepdiff: Single-stage b-rep diffusion model},
  author={Lee, Mingi and Zhang, Dongsu and Jambon, Cl{\'e}ment and Kim, Young Min},
  booktitle={Proceedings of the Special Interest Group on Computer Graphics and Interactive Techniques Conference Conference Papers},
  pages={1--11},
  year={2025}
}

@inproceedings{guo2025brepgiff,
  title={BrepGiff: Lightweight Generation of Complex B-rep with 3D GAT Diffusion},
  author={Guo, Hao and Huang, Xiaoshui and Bai, Yunpeng and Gan, Hongping and Shi, Yilei and others},
  booktitle={Proceedings of the Computer Vision and Pattern Recognition Conference},
  pages={26587--26596},
  year={2025}
}

@article{liu2025hola,
  title={Hola: B-rep generation using a holistic latent representation},
  author={Liu, Yilin and Xu, Duoteng and Yu, Xingyao and Xu, Xiang and Cohen-Or, Daniel and Zhang, Hao and Huang, Hui},
  journal={ACM Transactions on Graphics (TOG)},
  volume={44},
  number={4},
  pages={1--25},
  year={2025},
  publisher={ACM New York, NY, USA}
}

@inproceedings{xu2025autobrep,
  title={AutoBrep: Autoregressive B-Rep Generation with Unified Topology and Geometry},
  author={Xu, Xiang and Jayaraman, Pradeep and Lambourne, Joseph and Liu, Yilin and Malpure, Durvesh and Meltzer, Pete},
  booktitle={Proceedings of the SIGGRAPH Asia 2025 Conference Papers},
  pages={1--12},
  year={2025}
}

@article{li2025brepgpt,
  title={BrepGPT: Autoregressive B-rep Generation with Voronoi Half-Patch},
  author={Li, Pu and Zhang, Wenhao and Quan, Weize and Zhang, Biao and Wonka, Peter and Yan, Dongming},
  journal={ACM Transactions on Graphics (TOG)},
  volume={44},
  number={6},
  pages={1--18},
  year={2025},
  publisher={ACM New York, NY, USA}
}

@article{vaswani2017attention,
  title={Attention is all you need},
  author={Vaswani, Ashish and Shazeer, Noam and Parmar, Niki and Uszkoreit, Jakob and Jones, Llion and Gomez, Aidan N and Kaiser, {\L}ukasz and Polosukhin, Illia},
  journal={Advances in neural information processing systems},
  volume={30},
  year={2017}
}

@article{lai2025graphbrep,
  title={GraphBrep: Learning B-Rep in Graph Structure for Efficient CAD Generation},
  author={Lai, Weilin and Xu, Tie and Wang, Hu},
  journal={arXiv preprint arXiv:2507.04765},
  year={2025}
}

@article{ho2020denoising,
  title={Denoising diffusion probabilistic models},
  author={Ho, Jonathan and Jain, Ajay and Abbeel, Pieter},
  journal={Advances in neural information processing systems},
  volume={33},
  pages={6840--6851},
  year={2020}
}

@inproceedings{ma2025clr,
  title={Clr-wire: Towards continuous latent representations for 3d curve wireframe generation},
  author={Ma, Xueqi and Liu, Yilin and Gao, Tianlong and Huang, Qirui and Huang, Hui},
  booktitle={Proceedings of the Special Interest Group on Computer Graphics and Interactive Techniques Conference Conference Papers},
  pages={1--11},
  year={2025}
}

@inproceedings{li2025cad,
  title={CAD-Llama: leveraging large language models for computer-aided design parametric 3D model generation},
  author={Li, Jiahao and Ma, Weijian and Li, Xueyang and Lou, Yunzhong and Zhou, Guichun and Zhou, Xiangdong},
  booktitle={Proceedings of the Computer Vision and Pattern Recognition Conference},
  pages={18563--18573},
  year={2025}
}

@article{yin2025rlcad,
  title={Rlcad: Reinforcement learning training gym for revolution involved cad command sequence generation},
  author={Yin, Xiaolong and Lu, Xingyu and Shen, Jiahang and Ni, Jingzhe and Li, Hailong and Tong, Ruofeng and Tang, Min and Du, Peng},
  journal={arXiv preprint arXiv:2503.18549},
  year={2025}
}

@article{yuan2025cad,
  title={Cad-editor: A locate-then-infill framework with automated training data synthesis for text-based cad editing},
  author={Yuan, Yu and Sun, Shizhao and Liu, Qi and Bian, Jiang},
  journal={arXiv preprint arXiv:2502.03997},
  year={2025}
}

@article{zhang2024flexcad,
  title={Flexcad: Unified and versatile controllable cad generation with fine-tuned large language models},
  author={Zhang, Zhanwei and Sun, Shizhao and Wang, Wenxiao and Cai, Deng and Bian, Jiang},
  journal={arXiv preprint arXiv:2411.05823},
  year={2024}
}

@inproceedings{deng2025mamtiff,
  title={MamTiff-CAD: Multi-Scale Latent Diffusion with Mamba+ for Complex Parametric Sequence},
  author={Deng, Liyuan and Bai, Yunpeng and Dai, Yongkang and Huang, Xiaoshui and Gan, Hongping and Huang, Dongshuo and Jiacheng, Hao and Shi, Yilei},
  booktitle={Proceedings of the IEEE/CVF International Conference on Computer Vision},
  pages={10517--10526},
  year={2025}
}

@inproceedings{koch2019abc,
  title={Abc: A big cad model dataset for geometric deep learning},
  author={Koch, Sebastian and Matveev, Albert and Jiang, Zhongshi and Williams, Francis and Artemov, Alexey and Burnaev, Evgeny and Alexa, Marc and Zorin, Denis and Panozzo, Daniele},
  booktitle={Proceedings of the IEEE/CVF conference on computer vision and pattern recognition},
  pages={9601--9611},
  year={2019}
}

@inproceedings{siddiqui2024meshgpt,
  title={Meshgpt: Generating triangle meshes with decoder-only transformers},
  author={Siddiqui, Yawar and Alliegro, Antonio and Artemov, Alexey and Tommasi, Tatiana and Sirigatti, Daniele and Rosov, Vladislav and Dai, Angela and Nie{\ss}ner, Matthias},
  booktitle={Proceedings of the IEEE/CVF conference on computer vision and pattern recognition},
  pages={19615--19625},
  year={2024}
}

@article{hao2024meshtron,
  title={Meshtron: High-fidelity, artist-like 3d mesh generation at scale},
  author={Hao, Zekun and Romero, David W and Lin, Tsung-Yi and Liu, Ming-Yu},
  journal={arXiv preprint arXiv:2412.09548},
  year={2024}
}

@inproceedings{ma2024generating,
  title={Generating 3D House Wireframes with Semantics},
  author={Ma, Xueqi and Liu, Yilin and Zhou, Wenjun and Wang, Ruowei and Huang, Hui},
  booktitle={European Conference on Computer Vision},
  pages={223--240},
  year={2024},
  organization={Springer}
}

@inproceedings{xue2024neat,
  title={Neat: Distilling 3d wireframes from neural attraction fields},
  author={Xue, Nan and Tan, Bin and Xiao, Yuxi and Dong, Liang and Xia, Gui-Song and Wu, Tianfu and Shen, Yujun},
  booktitle={Proceedings of the IEEE/CVF Conference on Computer Vision and Pattern Recognition},
  pages={19968--19977},
  year={2024}
}

@inproceedings{zhu2023nerve,
  title={Nerve: Neural volumetric edges for parametric curve extraction from point cloud},
  author={Zhu, Xiangyu and Du, Dong and Chen, Weikai and Zhao, Zhiyou and Nie, Yinyu and Han, Xiaoguang},
  booktitle={Proceedings of the IEEE/CVF Conference on Computer Vision and Pattern Recognition},
  pages={13601--13610},
  year={2023}
}

@inproceedings{huang2024pbwr,
  title={PBWR: Parametric-building-wireframe reconstruction from aerial LiDAR point clouds},
  author={Huang, Shangfeng and Wang, Ruisheng and Guo, Bo and Yang, Hongxin},
  booktitle={Proceedings of the IEEE/CVF Conference on Computer Vision and Pattern Recognition},
  pages={27778--27787},
  year={2024}
}

@inproceedings{ma20223d,
  title={How-3d: Holistic 3d wireframe perception from a single image},
  author={Ma, Wenchao and Tan, Bin and Xue, Nan and Wu, Tianfu and Zheng, Xianwei and Xia, Gui-Song},
  booktitle={2022 International Conference on 3D Vision (3DV)},
  pages={596--605},
  year={2022},
  organization={IEEE}
}

@inproceedings{pautrat2023deeplsd,
  title={Deeplsd: Line segment detection and refinement with deep image gradients},
  author={Pautrat, R{\'e}mi and Barath, Daniel and Larsson, Viktor and Oswald, Martin R and Pollefeys, Marc},
  booktitle={Proceedings of the IEEE/CVF Conference on Computer Vision and Pattern Recognition},
  pages={17327--17336},
  year={2023}
}

@article{liu2021pc2wf,
  title={Pc2wf: 3d wireframe reconstruction from raw point clouds},
  author={Liu, Yujia and D'Aronco, Stefano and Schindler, Konrad and Wegner, Jan Dirk},
  journal={arXiv preprint arXiv:2103.02766},
  year={2021}
}

@article{tan2022coarse,
  title={Coarse-to-fine pipeline for 3D wireframe reconstruction from point cloud},
  author={Tan, Xuefeng and Zhang, Dejun and Tian, Long and Wu, Yiqi and Chen, Yilin},
  journal={Computers \& Graphics},
  volume={106},
  pages={288--298},
  year={2022},
  publisher={Elsevier}
}

@inproceedings{wang2022neural,
  title={Neural face identification in a 2d wireframe projection of a manifold object},
  author={Wang, Kehan and Zheng, Jia and Zhou, Zihan},
  booktitle={Proceedings of the IEEE/CVF Conference on Computer Vision and Pattern Recognition},
  pages={1622--1631},
  year={2022}
}

@inproceedings{ye2023nef,
  title={Nef: Neural edge fields for 3d parametric curve reconstruction from multi-view images},
  author={Ye, Yunfan and Yi, Renjiao and Gao, Zhirui and Zhu, Chenyang and Cai, Zhiping and Xu, Kai},
  booktitle={Proceedings of the IEEE/CVF Conference on Computer Vision and Pattern Recognition},
  pages={8486--8495},
  year={2023}
}

@article{guo2022complexgen,
  title={Complexgen: Cad reconstruction by b-rep chain complex generation},
  author={Guo, Haoxiang and Liu, Shilin and Pan, Hao and Liu, Yang and Tong, Xin and Guo, Baining},
  journal={ACM Transactions on Graphics (TOG)},
  volume={41},
  number={4},
  pages={1--18},
  year={2022},
  publisher={ACM New York, NY, USA}
}

@inproceedings{cherenkova2023sepicnet,
  title={Sepicnet: Sharp edges recovery by parametric inference of curves in 3d shapes},
  author={Cherenkova, Kseniya and Dupont, Elona and Kacem, Anis and Arzhannikov, Ilya and Gusev, Gleb and Aouada, Djamila},
  booktitle={Proceedings of the IEEE/CVF Conference on Computer Vision and Pattern Recognition},
  pages={2727--2735},
  year={2023}
}

@inproceedings{matveev20213d,
  title={3d parametric wireframe extraction based on distance fields},
  author={Matveev, Albert and Artemov, Alexey and Zorin, Denis and Burnaev, Evgeny},
  booktitle={Proceedings of the 2021 4th International Conference on Artificial Intelligence and Pattern Recognition},
  pages={316--322},
  year={2021}
}

@article{matveev2022def,
  title={Def: Deep estimation of sharp geometric features in 3d shapes},
  author={Matveev, Albert and Rakhimov, Ruslan and Artemov, Alexey and Bobrovskikh, Gleb and Egiazarian, Vage and Bogomolov, Emil and Panozzo, Daniele and Zorin, Denis and Burnaev, Evgeny},
  journal={ACM Transactions on Graphics},
  volume={41},
  number={4},
  year={2022}
}

@article{vinyals2015pointer,
  title={Pointer networks},
  author={Vinyals, Oriol and Fortunato, Meire and Jaitly, Navdeep},
  journal={Advances in neural information processing systems},
  volume={28},
  year={2015}
}

@article{song2020denoising,
  title={Denoising diffusion implicit models},
  author={Song, Jiaming and Meng, Chenlin and Ermon, Stefano},
  journal={arXiv preprint arXiv:2010.02502},
  year={2020}
}

@article{brown2020language,
  title={Language models are few-shot learners},
  author={Brown, Tom and Mann, Benjamin and Ryder, Nick and Subbiah, Melanie and Kaplan, Jared D and Dhariwal, Prafulla and Neelakantan, Arvind and Shyam, Pranav and Sastry, Girish and Askell, Amanda and others},
  journal={Advances in neural information processing systems},
  volume={33},
  pages={1877--1901},
  year={2020}
}

@inproceedings{chen2020generative,
  title={Generative pretraining from pixels},
  author={Chen, Mark and Radford, Alec and Child, Rewon and Wu, Jeffrey and Jun, Heewoo and Luan, David and Sutskever, Ilya},
  booktitle={International conference on machine learning},
  pages={1691--1703},
  year={2020},
  organization={PMLR}
}

@inproceedings{ramesh2021zero,
  title={Zero-shot text-to-image generation},
  author={Ramesh, Aditya and Pavlov, Mikhail and Goh, Gabriel and Gray, Scott and Voss, Chelsea and Radford, Alec and Chen, Mark and Sutskever, Ilya},
  booktitle={International conference on machine learning},
  pages={8821--8831},
  year={2021},
  organization={Pmlr}
}

@inproceedings{jayaraman2021uv,
  title={Uv-net: Learning from boundary representations},
  author={Jayaraman, Pradeep Kumar and Sanghi, Aditya and Lambourne, Joseph G and Willis, Karl DD and Davies, Thomas and Shayani, Hooman and Morris, Nigel},
  booktitle={Proceedings of the IEEE/CVF conference on computer vision and pattern recognition},
  pages={11703--11712},
  year={2021}
}

@inproceedings{lee2022autoregressive,
  title={Autoregressive image generation using residual quantization},
  author={Lee, Doyup and Kim, Chiheon and Kim, Saehoon and Cho, Minsu and Han, Wook-Shin},
  booktitle={Proceedings of the IEEE/CVF conference on computer vision and pattern recognition},
  pages={11523--11532},
  year={2022}
}

@misc{pyOCCT,
   author = {Trevor Laughlin},
   year = {2020},
   note = {https://github.com/trelau/pyOCCT},
   title = {pyOCCT -- Python bindings for OpenCASCADE via pybind11}
}

@inproceedings{dauphin2017language,
  title={Language modeling with gated convolutional networks},
  author={Dauphin, Yann N and Fan, Angela and Auli, Michael and Grangier, David},
  booktitle={International conference on machine learning},
  pages={933--941},
  year={2017},
  organization={PMLR}
}

@article{zhao2023michelangelo,
  title={Michelangelo: Conditional 3d shape generation based on shape-image-text aligned latent representation},
  author={Zhao, Zibo and Liu, Wen and Chen, Xin and Zeng, Xianfang and Wang, Rui and Cheng, Pei and Fu, Bin and Chen, Tao and Yu, Gang and Gao, Shenghua},
  journal={Advances in neural information processing systems},
  volume={36},
  pages={73969--73982},
  year={2023}
}

@article{xu2022rfeps,
  title={Rfeps: Reconstructing feature-line equipped polygonal surface},
  author={Xu, Rui and Wang, Zixiong and Dou, Zhiyang and Zong, Chen and Xin, Shiqing and Jiang, Mingyan and Ju, Tao and Tu, Changhe},
  journal={ACM Transactions on Graphics (TOG)},
  volume={41},
  number={6},
  pages={1--15},
  year={2022},
  publisher={ACM New York, NY, USA}
}

@article{contero2023quantitative,
  title={A quantitative analysis of parametric CAD model complexity and its relationship to perceived modeling complexity},
  author={Contero Gonz{\'a}lez, Manuel Roberto and P{\'e}rez L{\'o}pez, David Clemente and Camba, JD and others},
  year={2023},
  publisher={Elsevier}
}

\clearpage

\begin{figure*}[h]
    \centering
    \includegraphics[width=0.85\textwidth]{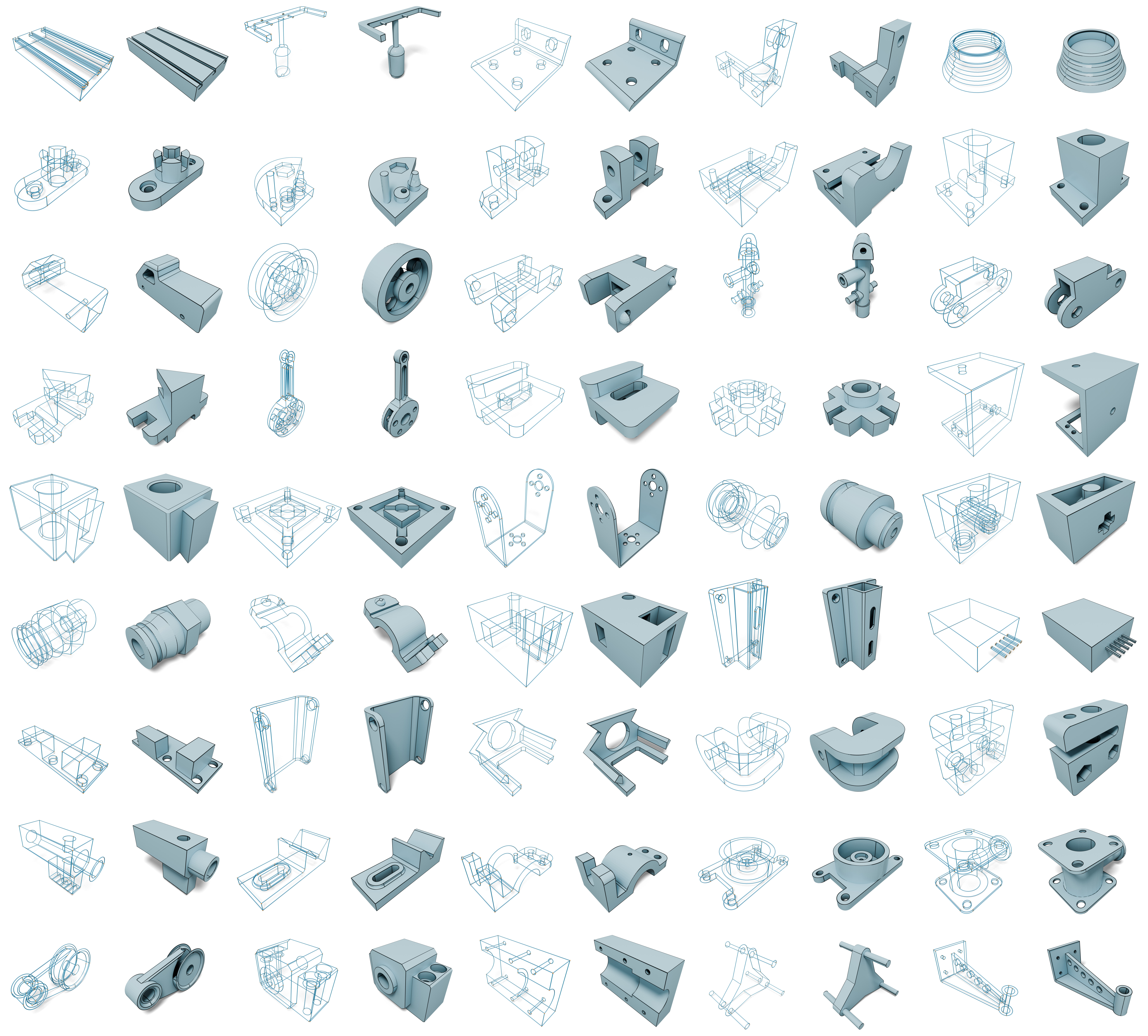}
    \caption{Gallery of unconditionally synthesized B-reps. We showcase a diverse range of synthesized CAD models, presenting both the generated wireframe scaffolds and their corresponding instantiated B-rep solids.}
    \label{fig: uncond_more}
\end{figure*}

\clearpage

\appendix

\section{Framework Details}

\subsection{Training and Inference}
In the wireframe generation stage, the model is trained autoregressively using cross-entropy loss with teacher forcing. Although the B-rep hierarchy imposes strict transition constraints (e.g., a $\langle \texttt{loop\_start} \rangle$ must precede a vertex), we compute the loss over the full vocabulary without transition masks. This soft supervision empirically leads to more stable optimization than hard-masked training. During inference, sequences are generated starting from $\langle \texttt{SOS} \rangle$ and terminated at $\langle \texttt{EOS} \rangle$ via nucleus sampling ($p=0.95$). A constrained sampling mask is applied based on the hierarchical state to restrict successors to valid tokens. The generated face-loops are finally processed by the merging scheme to recover the complete B-rep wireframe.

In the surface geometry refinement stage, the face VAE is kept fixed, and the Transformer-based refinement network is trained end-to-end using the loss in Eq.~\ref{eq: loss}. During inference, we first construct face geometry priors from the generated wireframe. These priors, together with the corresponding wireframe features, are then fed into the refinement network to predict surface geometries. The refined faces are subsequently assembled into the final B-rep model.

\subsection{Decoding of Complex Curves}
To reconstruct the geometric profiles of complex curves (e.g., B-splines), we adopt a decode-then-scale pipeline following~\cite{ma2025clr}. For each curve, a pretrained VQ-VAE decodes the 12-token sequence into 32 discrete points in a canonical coordinate space $[-1,1]$. The decoded curve is then transformed and scaled to align its endpoints with the predicted vertex coordinates in the global B-rep space.

As noted in prior work, this scaling process is sensitive to the distance between endpoints: as the edge length approaches zero, small perturbations in vertex positions can induce large errors in the scale factor. In our framework, numerical stability is ensured by the $1024^3$ vertex quantization grid (Sec.~\ref{sec: quant}), which enforces a minimum representable edge length of $2/1024 \approx 0.00195$. This lower bound prevents the scale factor from becoming ill-conditioned. Dataset statistics further show that edges near this resolution limit are rare, accounting for only $0.37\%$ of all edges. For these cases, the $10$-bit quantization still provides sufficient coordinate precision to maintain stable scaling. 

\section{Extended Analysis and Discussions}

\subsection{Curve Type Distributions}
To further evaluate the fidelity of our generative model, we analyze the distribution of curve types across the training dataset and the generated samples. As summarized in Tab.~\ref{tab:curve_dist}, our model closely mirrors the underlying data statistics. Specifically, the generated results maintain a balanced proportion of canonical primitives (Lines and Arcs) and complex B-spline curves, with only marginal deviations from the ground-truth distribution. This alignment suggests that BrepForge effectively learns the structural preferences of CAD models without suffering from mode collapse toward simpler geometries.

\begin{table}[tb]
\centering
\captionsetup{skip=4pt}
\setlength{\tabcolsep}{4pt}
\caption{Comparison of curve type distributions between the training set and generated samples.}
\label{tab:curve_dist}
\begin{tabular}{lccc}
\toprule
Category & Training Set ($\%$) & Generated Set ($\%$) & Difference($\%$) \\ 
\midrule
Line & 45.9    & 48.7  & 2.8       \\
Arc   & 30.5   & 32.1  & 1.6  \\
Complex & 23.6   & 19.2  & 4.4  \\ 
\bottomrule
\end{tabular}
\end{table}

\subsection{Analysis of Complex Curve Tokenization}
\begin{table}[tb]
\centering
\captionsetup{skip=4pt}
\caption{Reconstruction error (Chamfer Distance) for complex curves with varying token counts.}
\label{tab:token_ablation}
\begin{tabular}{lccc}
\toprule
Number of Tokens & 12 & 8 & 4 \\ 
\midrule
Reconstruction Error & $2 \times 10^{-5}$ & $6 \times 10^{-4}$ & $1 \times 10^{-2}$ \\ 
\bottomrule
\end{tabular}
\end{table}

The design of our complex curve representation balances geometric fidelity and autoregressive complexity. We provide additional analysis to justify the use of 12 tokens per curve and the resulting sequence lengths.

\textbf{Reconstruction Fidelity vs. Token Count.} To determine the optimal number of tokens for complex curves, we conducted a sensitivity analysis on the RQ-VAE bottleneck. As shown in Tab.~\ref{tab:token_ablation}, reducing the token count leads to a significant increase in reconstruction error (measured by Chamfer Distance). Specifically, decreasing the count from 12 to 4 tokens results in a nearly three-order-of-magnitude drop in precision ($2 \times 10^{-5}$ vs. $1 \times 10^{-2}$). Given that CAD modeling demands high geometric precision for downstream applications, we prioritize 12 tokens to ensure that complex shapes are captured with minimal distortion.

\textbf{Sequence Length and Complexity.} 
Complex curves account for $23.6\%$ of edges in our dataset (Tab.~\ref{tab:curve_dist}). Although 12 tokens per curve may appear costly, they are only assigned to complex edges, while lines and arcs use more compact representations. In our dataset, the average token count per face is $47$. For a typical CAD model with $70$ faces, the total sequence length is approximately $3,300$ tokens. We implement an upper bound of $8,000$ tokens to accommodate high-complexity outliers, ensuring the model's robustness across diverse engineering designs.

\textbf{Inference Efficiency.} Despite the increased sequence length, our framework maintains efficient inference through KV-caching. On a single GPU, the generation speed is approximately $120$ tokens per second. Consequently, even a complex model with an $8,000$-token sequence can be synthesized in approximately one minute. This demonstrates that 12 tokens strike a practical balance between geometric precision and inference efficiency.

\subsection{Validity Analysis Across Model Complexity}
\label{app:validity_complexity}

To address the potential bias introduced by simpler geometries in B-rep datasets, we group our generated samples based on their face counts and report the validity rates for these increasingly complex subsets. As shown in Tab.~\ref{tab:validity_face_count}, BrepForge maintains a high degree of topological integrity even as the structural complexity scales. Specifically, our model achieves a validity rate of 66.4\% for solids with more than 20 faces, which only gracefully degrades to 57.6\% for highly complex models with over 60 faces. These results demonstrate that our framework's performance does not collapse as geometric and topological demands increase, confirming its effectiveness in synthesizing sophisticated engineering designs.
\begin{table}[tb]
\centering
\captionsetup{skip=4pt}
\caption{Validity rates of BrepForge grouped by face count complexity.}
\label{tab:validity_face_count}
\begin{tabular}{lccc}
\toprule
Complexity Tier & $>$ 20 Faces & $>$ 40 Faces & $>$ 60 Faces \\ 
\midrule
Validity Rate ($\%$) & 66.4   & 62.0    & 57.6    \\ 
\bottomrule
\end{tabular}
\end{table}

\subsection{Effectiveness of Hierarchical Structural Embedding}

\begin{table}[tb]
\centering
\captionsetup{skip=4pt}
\caption{Ablation study of design components in B-rep generation. Both MMD and JSD scores are multiplied by $10^2$. $\text{Ours}_{\text{abs}}$ replaces our hierarchical structural embedding with a standard absolute positional encoding. $\text{Ours}_{\text{BFS}}$ substitutes our face-aware serialization strategy with a global Breadth-First Search (BFS) traversal.}
\label{tab: other-ablation}
\setlength{\tabcolsep}{3pt}
\begin{tabular}{lccc|ccc|c}
\toprule
Method 
& COV & MMD & JSD & Novel & Unique & Valid & CC \\
& (\%$\uparrow$) & ($\downarrow$) & ($\downarrow$)
& (\%$\uparrow$) & (\%$\uparrow$) & (\%$\uparrow$) & ($\uparrow$) \\
\midrule
Ours$_{\text{abs}}$
& 69.73 & 1.61 & 1.24 & 99.6 & 96.5 & 65.6 & 40.3 \\
Ours$_{\text{BFS}}$
& 70.07 & 1.39 & 1.12 & 99.7 & 97.0 & 67.4 & 42.8 \\
\textbf{Ours}
& \textbf{72.65} & \textbf{1.24} & \textbf{0.92} & \textbf{99.8} & \textbf{97.3} & \textbf{70.9} & \textbf{47.2} \\
\bottomrule
\end{tabular}
\end{table}

To validate the effectiveness of our Hierarchical Structural Embedding, we conduct an ablation study against a standard absolute positional encoding scheme~\cite{vaswani2017attention}. In this variant, the hierarchical relationships among faces, loops, and edges are replaced by flat sequential positional encodings. As reported in Tab.~\ref{tab: other-ablation} (denoted as Ours$_{\text{abs}}$), our full model consistently outperforms this baseline across all metrics. In particular, the validity drops from $70.9\%$ to $65.6\%$, accompanied by a notable decrease in Cyclomatic Complexity (CC), indicating degraded structural richness. This gap arises because B-rep models exhibit inherently hierarchical and nested topology rather than simple linear sequences. By explicitly encoding such structure, our hierarchical embedding introduces a strong topological inductive bias, enabling the model to better capture connectivity and spatial constraints in B-rep manifolds. 

\subsection{Impact of Serialization Strategies}
To evaluate the effectiveness of our face-aware serialization, we compare it with a global Breadth-First Search (BFS) traversal strategy adopted in AutoBrep\cite{xu2025autobrep}. While BFS shortens sequence length, it leads to fragmented geometric context and discontinuous topological connections. This places an additional burden on the Transformer, forcing it to model complex traversal patterns rather than focusing on intrinsic geometric structures. As shown in Tab.~\ref{tab: other-ablation}, our face-aware strategy consistently outperforms the BFS variant (denoted as Ours$_{\text{BFS}}$) across all metrics. By providing a more localized and geometrically coherent representation, our method enables the model to better capture underlying CAD design patterns, resulting in more complex and topologically consistent generations.

\subsection{Robustness to Point Cloud Noise}

To assess the robustness of our point-cloud-conditioned B-rep reconstruction, we evaluate the model's performance under varying levels of Gaussian noise. We jitter the input point clouds with zero-mean Gaussian noise at standard deviations of $\sigma \in \{0.002, 0.005, 0.01\}$, in addition to a baseline with clean points ($\sigma=0$). As reported in Tab.~\ref{tab: noise}, BrepForge exhibits strong resilience to moderate noise. While geometric accuracy (measured by CD and EMD) expectedly decreases as noise increases, the F-score remains high ($>0.85$) for noise levels up to $\sigma=0.005$. Even under significant noise ($\sigma=0.01$), the model successfully maintains the structural integrity of the generated B-reps. These results confirm that our framework is capable of reconstructing high-quality CAD models from imperfect, real-world sensor data.

\begin{table}[tb]
\centering
\captionsetup{skip=4pt}
\caption{Quantitative evaluation of point cloud-conditioned B-rep reconstruction under varying noise levels. Both CD and EMD scores are multiplied by $10^2$.}
\label{tab: noise}
\begin{tabular}{lccc}
\toprule
Noise Level & CD ($\downarrow$) & EMD ($\downarrow$) & F-score ($\uparrow$) \\
\midrule
Clean ($\sigma=0$) & 0.87 & 2.19 & 0.92 \\
$\sigma=0.002$ & 0.92 & 2.31 & 0.90 \\
$\sigma=0.005$ & 1.26 & 2.64 & 0.85 \\
$\sigma=0.01$ & 2.01 & 3.37 & 0.77 \\
\bottomrule
\end{tabular}
\end{table}

\subsection{Limitations and Future Work}
\label{sec:limitations}

\begin{figure}[tb] 
    \centering
    \begin{subfigure}[b]{0.24\linewidth}
        \centering
        \includegraphics[width=\linewidth]{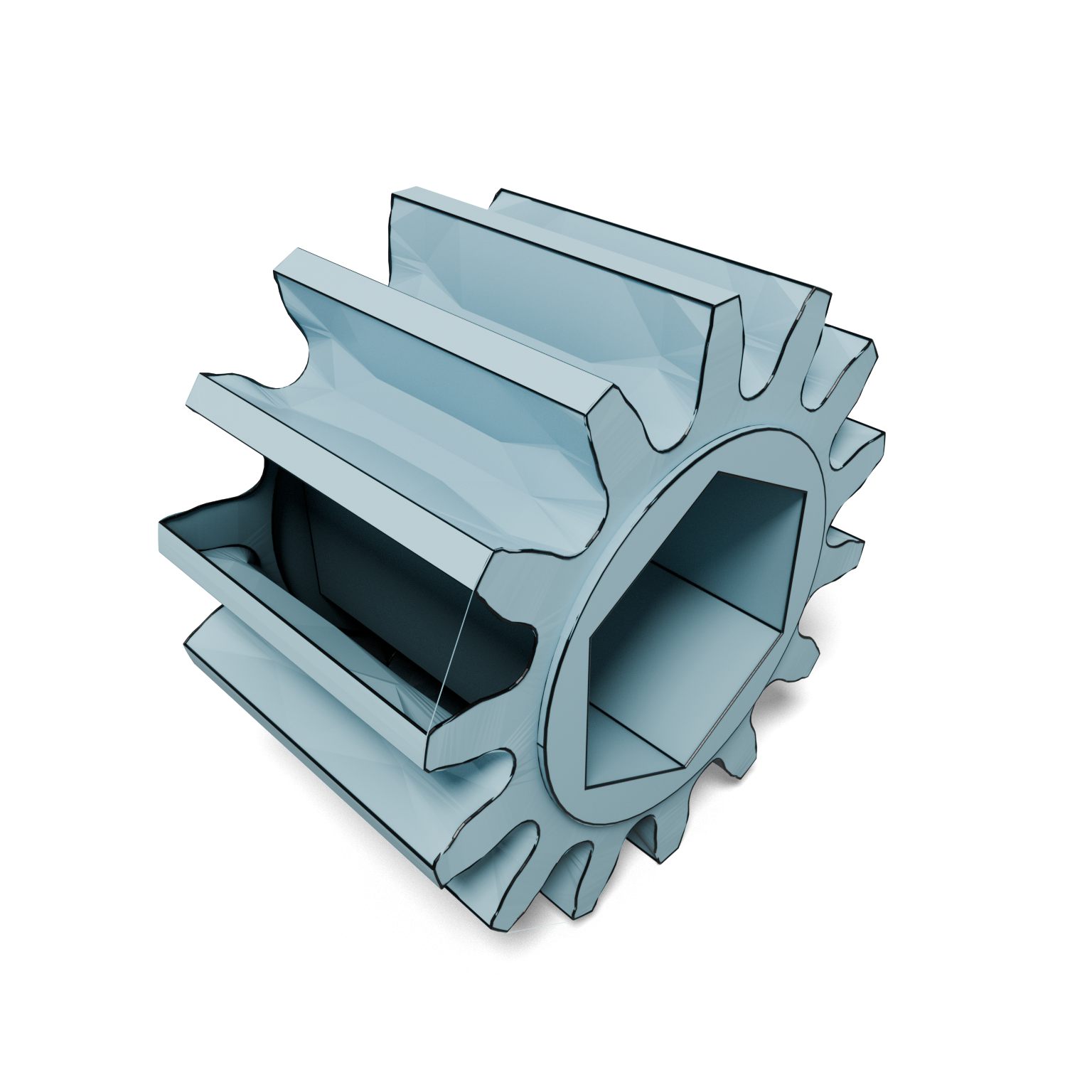}
        \label{fig:sub1}
    \end{subfigure}
    \hfill 
    \begin{subfigure}[b]{0.24\linewidth}
        \centering
        \includegraphics[width=\linewidth]{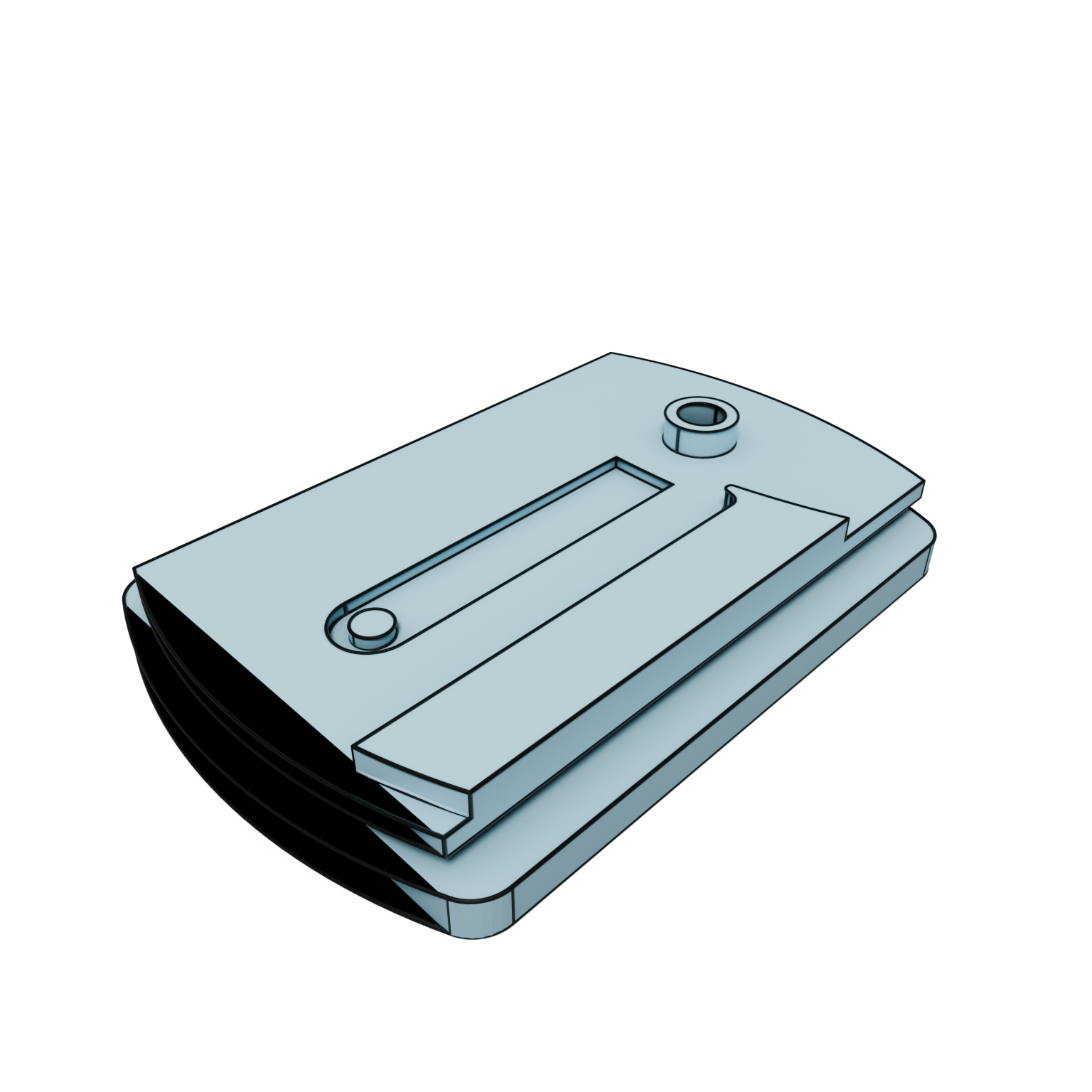}
        \label{fig:sub2}
    \end{subfigure}
    \hfill
    \begin{subfigure}[b]{0.24\linewidth}
        \centering
        \includegraphics[width=\linewidth]{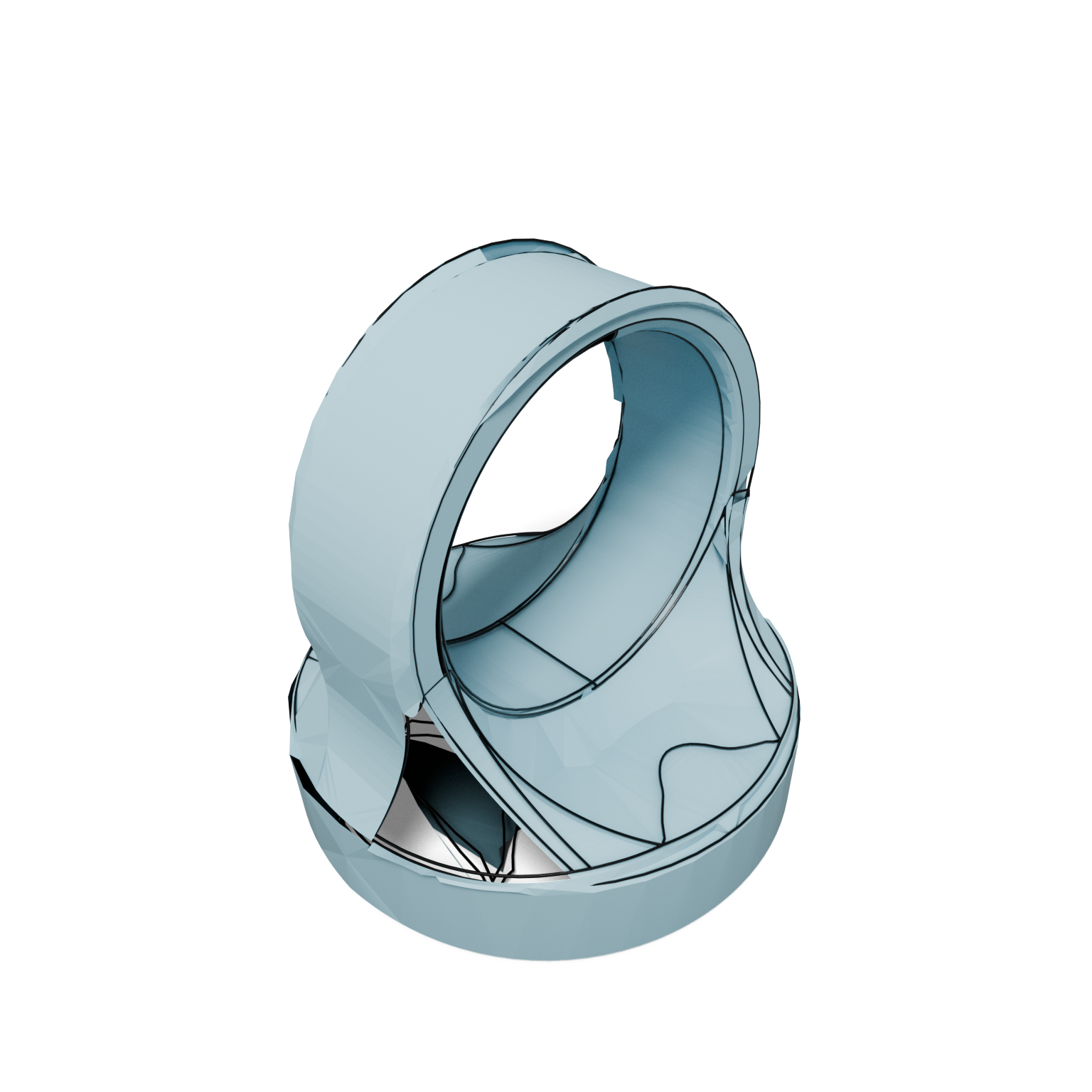}
        \label{fig:sub3}
    \end{subfigure}
    \hfill
    \begin{subfigure}[b]{0.24\linewidth}
        \centering
        \includegraphics[width=\linewidth]{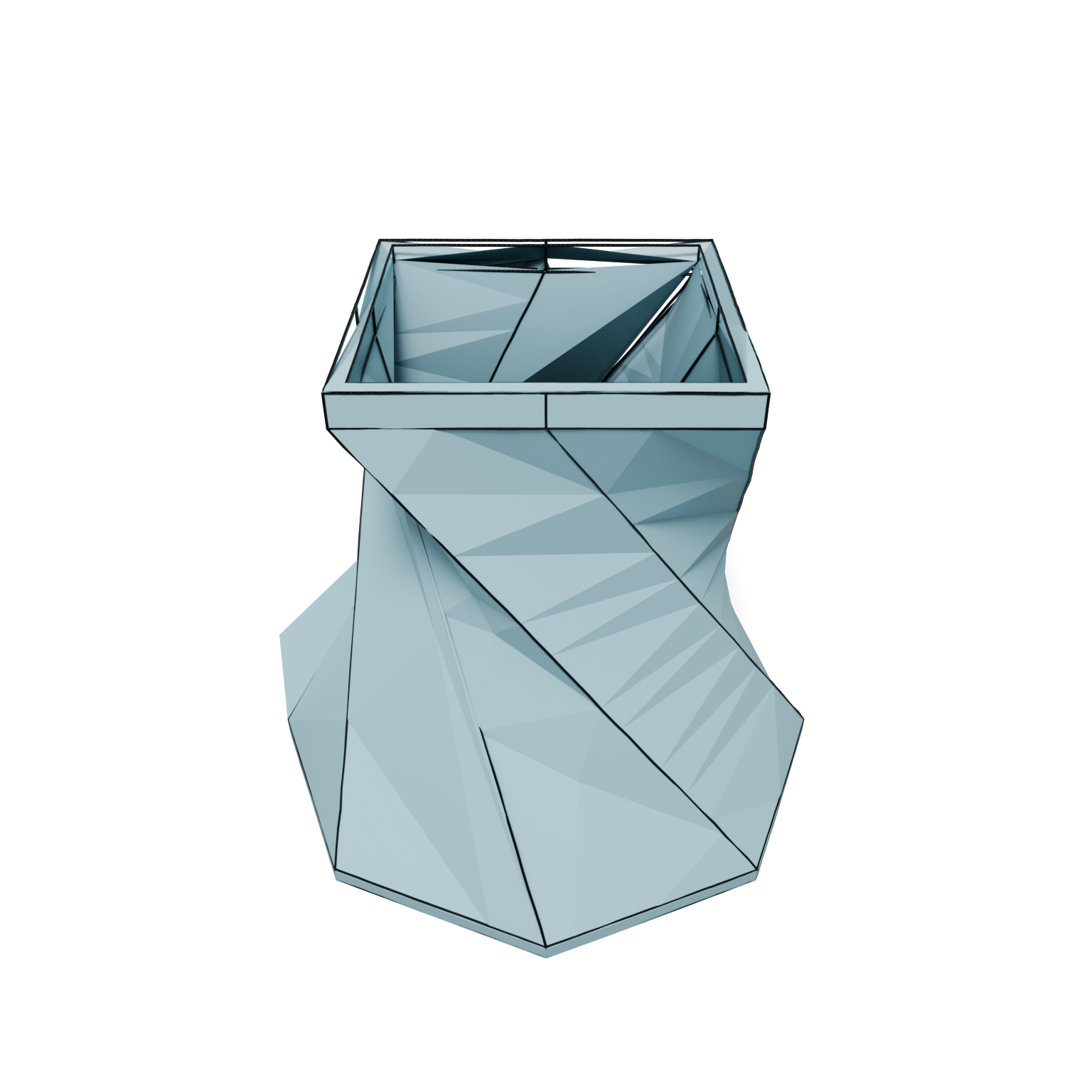}
        \label{fig:sub4}
    \end{subfigure}
    \captionsetup{skip=1pt}
    \caption{Representative failure cases generated by our method. These examples primarily demonstrate issues in topological recovery, such as unmerged shared edges, or geometric imperfections like tiny gaps and surface self-intersections in complex geometries.}
    \label{fig: bad}
\end{figure}

Despite promising results, our framework still has several limitations that suggest directions for future work.

\textbf{Robustness of Shared Entity Recovery.} Our face-aware serialization recovers global topology via a post-processing, distance-based merging step. While generally effective, this procedure can be sensitive to extreme geometric proximity or numerical precision, potentially missing shared edges or incorrectly merging distinct ones. Future work will explore incorporating shared-edge constraints into training or replacing the heuristic merging with a learned topological matching module.

\textbf{Scalability to High-Complexity Models.} Although BrepForge supports sequences up to 8,000 tokens and models with up to 70 faces, scalability remains limited by the quadratic complexity of Transformer attention. We plan to investigate hierarchical or linear-complexity attention mechanisms to better handle more complex CAD assemblies.

\textbf{Geometric Integrity and Manifold Soundness.} Like prior B-rep generative models, BrepForge does not always guarantee watertight and manifold-valid outputs for highly intricate designs (see Fig.~\ref{fig: bad}). Complex surface interactions may introduce small gaps or self-intersections that violate B-rep constraints. Future work will explore integrating topology-aware losses to encourage geometric and physical consistency for downstream applications.

\end{document}